\documentclass[12pt]{article}
\usepackage{amsmath}
\usepackage{amsthm}
\usepackage{amsfonts}
\usepackage[dvips]{epsfig}
\usepackage{graphicx}
\usepackage{psfrag}
\usepackage{amssymb}
\usepackage{cancel}
\usepackage{axodraw4j}
\usepackage{lscape}
\usepackage{slashed}
\usepackage{color}

\newcommand{\beq}{\begin{equation}}
\newcommand{\eeq}{\end{equation}}

\newcommand{\rend}{{\rm end}}
\newcommand{\bk}{\mathbf{k}}

\newcommand{\bp}{\mathbf{p}}

\begin{document}

\begin{flushright}
{\tt KCL-PH-TH/2015-46}, {\tt LCTS/2015-33}, {\tt CERN-PH-TH/2015-239}  \\
{\tt ACT-10-15, MI-TH-1541} \\
{\tt UMN-TH-3511/15, FTPI-MINN-15/48} \\
\end{flushright}

\vspace{1cm}
\begin{center}
{\bf {\large Post-Inflationary Gravitino Production Revisited} 
}
\vspace {0.1in}
\end{center}

\vspace{0.05in}

\begin{center}{
{\bf John~Ellis}$^{a}$,
{\bf Marcos~A.~G.~Garcia}$^{b}$,
{\bf Dimitri~V.~Nanopoulos}$^{c}$, \\
\vspace {0.1in}
{\bf Keith~A.~Olive}$^{b}$ and
{\bf Marco~Peloso}$^{b}$
}
\end{center}

\begin{center}
{\em $^a$Theoretical Particle Physics and Cosmology Group, Department of
  Physics, King's~College~London, London WC2R 2LS, United Kingdom;\\
Theory Division, CERN, CH-1211 Geneva 23,
  Switzerland}\\[0.2cm]
  {\em $^b$William I. Fine Theoretical Physics Institute, School of Physics and Astronomy,\\
University of Minnesota, Minneapolis, MN 55455, USA}\\[0.2cm]
{\em $^c$George P. and Cynthia W. Mitchell Institute for Fundamental Physics and Astronomy,
Texas A\&M University, College Station, TX 77843, USA;\\
Astroparticle Physics Group, Houston Advanced Research Center (HARC), \\ Mitchell Campus, Woodlands, TX 77381, USA;\\
Academy of Athens, Division of Natural Sciences,
Athens 10679, Greece}\\
\end{center}

\bigskip
\bigskip
\centerline{\bf ABSTRACT}
\vspace{0.2in}

\noindent  
{\small We revisit gravitino production following inflation. As a first step, we review the standard calculation of
gravitino production in the thermal plasma formed at the end of post-inflationary reheating when the inflaton
has completely decayed. Next we consider gravitino production prior to the completion of reheating, 
assuming that the inflaton decay products thermalize instantaneously while they are still dilute. 
We then argue that instantaneous thermalization is in general
a good approximation, and also show that the contribution of non-thermal gravitino
production via the collisions of inflaton decay products prior to thermalization is relatively small. 
Our final estimate of the gravitino-to-entropy ratio is approximated well by a standard calculation of gravitino production in the post-inflationary thermal plasma
assuming total instantaneous decay and thermalization at a time $t \simeq 1.2/\Gamma_\phi$.
Finally, in light of our calculations, we consider potential
implications of upper limits on the gravitino abundance for models of inflation, with particular
attention to scenarios for inflaton decays in supersymmetric Starobinsky-like models.}

\vspace{0.2in}
\bigskip
\bigskip
\begin{flushleft}
December 2015
\end{flushleft}
\medskip
\noindent

\newpage

\section{Introduction}
~\\
Measurements by the Planck satellite~\cite{Planck} and ground-based experiments such as BICEP2/Keck Array~\cite{PBK,BK}
are probing the cosmic microwave background (CMB) ever more precisely, a new round of experiments
is on the way, and far more accurate experiments are being proposed for the future. The data are
already putting strong pressure on models of cosmological inflation, excluding many and constraining
severely the survivors. The most stringent constraints are generally those from the magnitude and tilt, $n_s$,
of the scalar perturbation spectrum, and the tensor-to-scalar ratio, $r$, with constraints on non-Gaussianities,
isocurvature perturbations, etc., being less powerful in the context of most slow-roll models \cite{reviews}. 
The predictions
of these models depend, in general, on the number of e-folds of inflation, $N_*$, which depends in turn of the
amount of reheating at the end of inflation \cite{reheat,EGNO5}. 

Comparisons of the Planck data with inflationary models typically
consider $40 < N_* < 60$, and characteristic model predictions for $n_s$, in particular, vary by amounts
comparable with the 68\% experimental range in $n_s$ as $N_*$ varies over this range. This means that the
experimental data are already starting to provide interesting constraints on $N_*$, and hence indirectly on
the amount of reheating. The latter depends, in turn, on the decay rate of the inflaton into relativistic particles. For
example, if the dominant inflaton decay is into two particles, reheating and hence the required number of
e-folds $N_*$ and the predictions for $n_s$ and $r$ depend on the two-body decay coupling $y$. 

Conversely, the experimental constraints on $n_s$ and $r$ can be used to constrain the coupling $y$ in the context of
any specific model. For example, in models whose predictions resemble those of the Starobinsky $R^2$ model~\cite{Staro},
a combination of the Planck, BICEP2/Keck Array and BAO data yields $N_* \gtrsim 50$ at the 68\% CL,
corresponding to $y \gtrsim 5\times 10^{-8}$ \cite{EGNO5}.

In supersymmetric models, cosmological and astrophysical constraints on the abundance of gravitinos produced
after inflation yield complementary restrictions on the amount of reheating, and hence $y$ 
\cite{weinberg,eln,nos,ehnos,kl,ekn,Juszkiewicz:gg,mmy,Kawasaki:1994af,Moroi:1995fs,enor,Giudice:1999am,bbb,kmy,stef,Pradler:2006qh,ps2,rs}.
These constraints may arise from considerations of the relic dark matter density due to gravitinos or their decay
products and/or from limits on late-decaying gravitinos imposed, e..g., by the success of Big-Bang
nucleosynthesis calculations \cite{bbn,ceflos175,Kawasaki:1994af,stef}. These upper limits on the produced gravitino abundance translate into
an upper bound on a two-body coupling of $y \lesssim 10^{-5}$ \cite{EGNO5}.

In view of the present and prospective future constraints on inflaton decay via CMB limits on $N_*$
and the competition (in supersymmetric models) with constraints from the gravitino abundance, in this paper
we revisit the issue of gravitino production following inflation.
One source of gravitinos that is well understood is production by particle collisions in the thermal plasma
that fills the Universe after reheating \cite{ekn,mmy,Kawasaki:1994af,enor,Giudice:1999am,bbb,Pradler:2006qh,ps2,rs}. However, gravitinos could also have been produced by particle
collisions before the reheating process was complete, either by collisions of relativistic inflaton decay products
before thermalization, or in any dilute thermal plasma formed by their collisions while inflaton decay was
continuing \cite{Giudice:1999am,rs}. 

We consider all these mechanisms in this paper,~\footnote{On the other hand, we disregard the possible additional gravitino quanta 
that could be produced if the inflaton experiences a strong non-perturbative decay at the onset of its oscillations, also known as preheating, since
non-perturbative gravitino production at preheating has been shown to be small~\cite{Kallosh:1999jj,Giudice:1999yt,Giudice:1999am,Kallosh:2000ve,Nilles:2001ry,Nilles:2001fg,Nilles:2001my}. 
In this scenario,  gravitinos may also be perturbatively produced by the non-thermal distributions formed at preheating. We disregard this effect since it is model-dependent,
and since the evolution of these distributions is not well understood \cite{Podolsky:2005bw}.} and compare the naive approximation of total instantaneous decay
at $t = c /\Gamma_\phi$, where $\Gamma_\phi$ is the inflaton decay rate and $c$ is some ${\cal O}(1)$ constant,
with the exact solution involving continuous decays producing the thermal bath. 
Whilst the common approximation neglects gravitino production at times $t < c / \Gamma_\phi$,
it also neglects the dilution of the gravitino abundance due to radiation produced at $t > c / \Gamma_\phi$.
Remarkably,  the approximate and exact solutions agree for the choice $c = 1.2$~\footnote{We note that
two choices of $c$ are commonly found in the literature, namely $c = 1$, which is relatively accurate,
and $c = 2/3$, corresponding to the condition $\Gamma_\phi  = H$, where $H$ is the Hubble 
parameter. As shown in~\cite{ps2,rs}, the latter choice yields a gravitino abundance that is too large by a factor $\sim \sqrt{3/2}$.}.
Using the production rate for gravitinos, we then derive an analytic expression for later cosmological evolution assuming continuous inflaton decays. 

We apply these results to recent inflationary models based on no-scale supergravity \cite{ENO6,ENO7,ENO8,EGNO2,EGNO4}. In particular, we focus on the phenomenological aspects
of Starobinsky-like models of inflation \cite{EGNO4}. Reheating in no-scale models does not
occur automatically \cite{ekoty}, but instead requires either an explicit coupling 
of the inflaton to matter \cite{ENO6,ENO8,EGNO4}, a coupling to moduli \cite{EGNO4}, or
a coupling to the gauge sector through the gauge kinetic function \cite{ekoty,EGNO4}. 
Direct decays of the inflaton to gravitinos may, in general, compete with the thermal production,
and both should be considered when setting limits on the couplings governing inflaton decay.

The standard calculation of gravitino production in the thermal plasma formed after reheating is
reviewed in Section~2. Then, in Section~3 we calculate gravitino production before the completion of
reheating, assuming that the inflaton decay products thermalize instantaneously. These are, in some sense,
opposite extremes for the treatment of gravitino production. We also provide an analytic solution
to the late time gravitino yield when the instantaneous decay approximation is dropped.  In Section~4 we analyze the degree to which
instantaneous thermalization is a good approximation, and also discuss non-thermal contributions to the
gravitino production rate prior to thermalization. In Section~5 we review the consequences for
inflaton couplings and compare with the constraints from Big Bang Nucleosynthesis, the relic cold dark
matter density, and the CMB constraints on the number of inflationary e-folds. In particular, we
derive constraints on the couplings governing inflaton decay in phenomenological
no-scale inflation models~\cite{EGNO4,EGNO5}.  Finally, in Section~6
we summarize our conclusions.

\section{Production in the Thermal Plasma after Reheating}\label{sec2}

Most calculations of gravitino production assume the instantaneous decay of the inflaton 
at $t = 1/\Gamma_\phi$ where $\Gamma_\phi$ is the inflaton decay rate. 
Since the Universe is dominated by the oscillations of the inflaton prior to decay, 
and these oscillations act as matter, this is equivalent to $\Gamma_\phi = 3 H/2$. 
It is then assumed that the decay products thermalize very rapidly \cite{Davidson:2000er}.
In this case, one can define a reheating temperature from the instantaneous conversion of 
the energy density in oscillations to that of radiation,
\beq
T_{\rm reh} = \left(\frac{30\rho_{\rm reh}}{\pi^2 g_{\rm reh}}\right)^{1/4}\,,
\label{Trehrhoreh}
\eeq
where $\rho_{\rm reh}$ denotes the energy density of radiation and $g_{\rm reh}$ denotes the effective number of 
degrees of freedom at the `moment' of reheating. Gravitinos are then produced by scattering processes in the thermalized 
radiation-filled Universe, after which they decay with the following decay rate into particles within the minimal
supersymmetric extension of the Standard Model (MSSM)~\cite{Moroi:1995fs},
\beq\label{grav_decay}
\Gamma_{3/2} = \frac{193}{384\pi}\frac{m_{3/2}^3}{M_P^2}\,.
\eeq
Here $M_P$ refers to the reduced Planck mass, $M_P=(8\pi G_N)^{-1/2}\simeq 2.4\times 10^{18}\,{\rm GeV}$. 
Since the interactions of gravitinos are very weak, they do not thermalize with the radiation background. 
Therefore, the Boltzmann equation for the gravitino number density $n_{3/2}$ can be written as 
\beq\label{grav_boltz}
\frac{dn_{3/2}}{dt} + 3Hn_{3/2} = \langle\sigma_{\rm tot}v_{\rm rel}\rangle n_{\rm rad}^2 - \frac{m_{3/2}}{\langle E_{3/2}\rangle}\Gamma_{3/2}n_{3/2} \, ,
\eeq
where $\langle\sigma_{\rm tot}v_{\rm rel}\rangle$ is the thermally-averaged gravitino production cross section,
$n_{\rm rad}={\zeta(3)}T^3/{\pi^2}$ is the number density of any single bosonic relativistic degree of freedom,
and $\langle E_{3/2}\rangle/m_{3/2}$ is the averaged Lorentz factor. Inverse-scattering terms are omitted,
because their contributions are unimportant at the reheating temperatures of relevance~\cite{Kawasaki:1994af}.
 
The thermally-averaged cross section for the Standard Model
$SU(3)_c\times SU(2)_L\times U(1)_Y$ gauge group was calculated in~\cite{bbb,Pradler:2006qh,rs}. 
Including contributions from $2\rightarrow2$ gauge scatterings, production via $1\rightarrow2$ decays allowed by thermal masses, 
and the effect of the top Yukawa coupling $y_t$, it can be parametrized as
\beq
\langle \sigma_{\rm tot}v_{\rm rel}\rangle  =  \langle \sigma_{\rm tot}v_{\rm rel}\rangle_{\rm gauge} + \langle \sigma_{\rm tot}v_{\rm rel}\rangle_{\rm top} 
\eeq
with
\beq
\langle \sigma_{\rm tot}v_{\rm rel}\rangle_{\rm top} =  1.29\,\frac{|y_t|^2}{M_P^2}\left[1+\frac{A_t^2}{3m_{3/2}^2}\right] \,,
\eeq
where $A_t$ is the top-quark supersymmetry-breaking trilinear coupling, and
\begin{eqnarray}
\langle \sigma_{\rm tot}v_{\rm rel}\rangle_{\rm gauge} & = & \sum_{i=1}^3 \frac{3\pi c_ig_i^2}{16 \zeta(3) M_P^2} \left[1+\frac{m_{\tilde{g}_i}^2}{3m_{3/2}^2}\right]
\ln\left(\frac{k_i}{g_i}\right) \nonumber \\
& & \!\!\!\!  \!\!\!\!  \!\!\!\!  \!\!\!\!  \!\!\!\! 
=  \frac{26.24}{M_P^2}  \left[\left(1+0.558\,\frac{m_{1/2}^2}{m_{3/2}^2}\right) - 0.011 \left(1+3.062\,\frac{m_{1/2}^2}{m_{3/2}^2}\right) \log\left(\frac{T_{\rm reh}}{10^{10}\,{\rm GeV}}\right)\right] \, ,
\label{ck}
\end{eqnarray}
where the $m_{\tilde{g}_i}$ are the gaugino masses and the constants $c_i,k_i$ depend on the gauge group, as shown in Table~\ref{table:gauge}.
We have obtained these values through a phenomenological fit to the result of \cite{rs} using the convenient parametrization of \cite{Pradler:2006qh}, 
under the assumption of a unified gauge coupling $\alpha=1/24$ and universal
gaugino masses $m_{1/2}$ at the scale $M_{\rm GUT} = 2\times 10^{16}\,$GeV. Table~\ref{table:gauge} 
differs from the result of \cite{Pradler:2006qh} in that it includes gravitino production via decays, 
leading to a production rate that is about twice larger. We have included in (\ref{ck}) the leading logarithmic corrections to the 
running of all the gauge couplings and gaugino masses, which at one-loop order are given by
\begin{align}
g_i(T)^2 &= \frac{g_i(M_{\rm GUT} )^2}{1-\dfrac{b_i}{8\pi^2}\,g_i(M_{\rm GUT} )^2\,\ln(T/M_{\rm GUT} )}\, , \qquad \left(
\begin{matrix}
b'\\b\\b_s
\end{matrix}\right) = \left(
\begin{matrix}
11\\1\\-3
\end{matrix}
\right)\,,\\
m_{\tilde{g}_i}(T)& = \left(\frac{g_i(T)}{g_i(M_{\rm GUT})}\right)^2\, m_{1/2}\,.
\end{align}
It is worth noting that the first term in the gaugino mass-dependent factors $(1+m_{\tilde{g}_i}^2/3m_{3/2}^2)$ corresponds to the production of the transversally polarized gravitino, while the second term is associated with the production of the longitudinal (Goldstino) component.
\begin{table}[t]
\centering
\begin{tabular}[c]{c c c c}
\hline \hline Gauge group & $g_i$ & $c_i$ & $k_i$\\ 
\hline $U(1)_Y$ & $g'$ & 9.90 & 1.469\\ 
 $SU(2)_L$ & $g$ & 20.77 & 2.071\\ 
 $SU(3)_c$ & $g_s$ & 43.34 & 3.041\\ 
\hline \hline
\end{tabular} 
\caption{\em The values of the constants $c_i$ and $k_i$ in the parameterization (\protect\ref{ck})
for the Standard Model gauge groups $U(1)_Y$, $SU(2)_L$, and $SU(3)_c$. The values are obtained from  
a phenomenological fit to the result of \cite{rs}, as explained in the text.
}
\label{table:gauge}
\end{table}

Immediately after reheating, the second term in the r.h.s. of (\ref{grav_boltz}) is negligible. In terms of the gravitino yield
\beq
Y_{3/2}\equiv \frac{n_{3/2}}{n_{\rm rad}}\,,
\eeq
the Boltzmann equation (\ref{grav_boltz}) can be rewritten in the form
\beq\label{boltz2}
\dot{Y}_{3/2} + 3\left(H+\frac{\dot{T}}{T}\right)Y_{3/2} = \langle\sigma_{\rm tot}v_{\rm rel}\rangle n_{\rm rad}\,.
\eeq
Under the assumption of entropy conservation, $gT^3a^3=$ const., where $a$ is the cosmological scale factor. Eq. (\ref{boltz2}) is equivalent to
\beq\label{boltz3}
\frac{dY_{3/2}}{dT} - \frac{d\ln g}{dT}\,Y_{3/2} = -\frac{\langle\sigma_{\rm tot}v_{\rm rel}\rangle n_{\rm rad}}{HT}\left[1+\frac{T}{3} \frac{d\ln g}{dT}\right]\,.
\eeq
Straightforward integration then yields
\beq
Y_{3/2}(T)=Y_{3/2}(T_{\rm reh})\,\frac{g(T)}{g(T_{\rm reh})} - g(T)\int_{T_{\rm reh}}^T  \frac{\langle\sigma_{\rm tot}v_{\rm rel}\rangle n_{\rm rad}(\tau)}{g(\tau)H(\tau)\,\tau}\left[1+\frac{\tau}{3} \frac{d\ln g(\tau)}{d\tau}\right]\,d\tau\,.
\eeq
Notice that this integration begins at $T_{\rm reh}$, consistent with the
assumption that inflaton decay and thermalization are instantaneous and simultaneous at $T_{\rm reh}$, and runs to lower $T$.
Assuming a vanishing abundance at $T_{\rm reh}$, and disregarding the weak dependence on temperature of the integrand in the r.h.s., integration from $T_{\rm reh}$ to $T\ll T_{\rm reh}$ yields
\beq\label{Yapp}
Y_{3/2}(T) \simeq \frac{\langle\sigma_{\rm tot}v_{\rm rel}\rangle n_{\rm rad}(T_{\rm reh})}{H(T_{\rm reh})}\times \frac{g(T)}{g(T_{\rm reh})}\,.
\eeq
Hence the final abundance of gravitinos is given by the ratio of the production rate ($\langle\sigma_{\rm tot}v_{\rm rel}\rangle n_{\rm rad}$) to the Hubble rate at reheating, 
diluted by subsequent particle annihilations and accounting for the ratio of numbers of degrees of freedom at $T$ to that at $T_{\rm reh}$.

\begin{figure}[h!]
\centering
    \psfrag{y}[b]{\LARGE{$Y_{3/2}$
    }}
    \psfrag{x}{\LARGE{$T$ (GeV)}}
    \scalebox{0.6}{\includegraphics{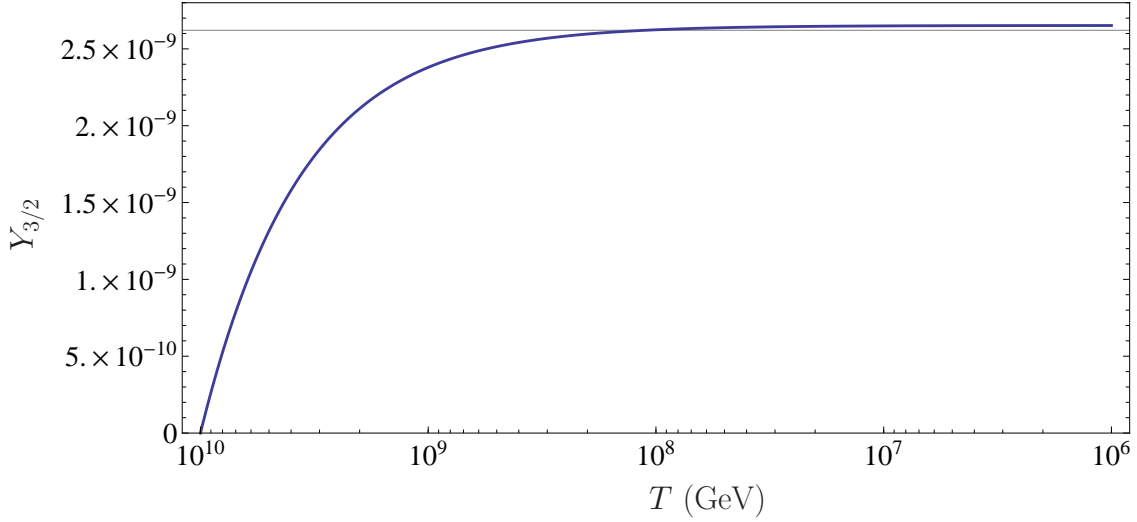}}
    \caption{\em The gravitino yield $Y_{3/2}$ for $T_{\rm reh}=10^{10}\,{\rm GeV}$ as a function of temperature during radiation domination following the completion of reheating. 
    The blue curve corresponds to the numerical integration of (\ref{boltz2}) under the simplifying assumption $g=915/4=$const., 
    disregarding the top Yukawa contribution and the $m_{1/2}$-dependent Goldstino component.
    The approximation (\ref{yield0}), not including the dilution factor $g(T)/g_{\rm reh}$, is displayed as the horizontal black line.
        }
    \label{fig:Yapp}
\end{figure}

During the radiation-dominated era, the cosmic time and temperature are related by
\beq
t=\sqrt{\frac{45}{2\pi^2g}}\frac{M_P}{T^2}\,.
\eeq
When the temperature of the Universe drops to $T\ll (\Gamma_{3/2}/\Gamma_{\phi})^{1/3}T_{\rm reh}$, 
where $\Gamma_{\phi}$ is the inflaton decay rate, the decay term in the Boltzmann equation (\ref{grav_boltz})
dominates over the scattering term. In this case, gravitinos have redshifted their momenta away, 
which implies that (\ref{grav_boltz}) may be rewritten as
\beq\label{bltzdec}
\dot{Y}_{3/2}=-\Gamma_{3/2}Y_{3/2}\,.
\eeq
Under the assumption that $m_{3/2} \ll 10^{13} ~ {\rm GeV} \times (T_{\rm reh}/10^{10} {\rm GeV})$,
the approximation (\ref{Yapp}) may be taken as an initial condition for (\ref{bltzdec}). Denoting $g_{\rm reh}=g(T_{\rm reh})$, the gauge contribution to the gravitino abundance can finally be written as
\begin{align}
Y_{3/2}(T)&\simeq \frac{\langle\sigma_{\rm tot}v_{\rm rel}\rangle n_{\rm rad}(T_{\rm reh})}{H(T_{\rm reh})}\times \frac{g(T)}{g_{\rm reh}} \times e^{-\Gamma_{3/2}t}\,,\nonumber \\ 
&\simeq 7.40\times 10^{-10}e^{-\Gamma_{3/2}t}\,\frac{g(T)}{g_{\rm reh}^{3/2}}\left(\frac{T_{\rm reh}}{10^{10}\,{\rm GeV}}\right) \sum_{i=1}^3 c_i\, g_i(T_{\rm reh})^2 \left(1+\frac{m_{\tilde{g}_i}(T_{\rm reh})^2}{3m_{3/2}^2}\right) \ln\left(\frac{k_i}{g_i(T_{\rm reh})}\right) \nonumber \\
 &\simeq 3.96\times 10^{-8}e^{-\Gamma_{3/2}t} \;\frac{g(T)}{g_{\rm reh}^{3/2}} \left(\frac{T_{\rm reh}}{10^{10}\,{\rm GeV}}\right) \nonumber\\
 &\qquad \times \left[\left(1+0.558\,\frac{m_{1/2}^2}{m_{3/2}^2}\right) - 0.011 \left(1+3.062\,\frac{m_{1/2}^2}{m_{3/2}^2}\right) \ln\left(\frac{T_{\rm reh}}{10^{10}\,{\rm GeV}}\right)\right]\,,
 \label{yield0}
\end{align}
or
\begin{align}
Y_{3/2}(T)&\simeq  4.48\times 10^{-11}e^{-\Gamma_{3/2}t} \left(\frac{T_{\rm reh}}{10^{10}\,{\rm GeV}}\right) \nonumber\\
 &\qquad \times \left[\left(1+0.558\,\frac{m_{1/2}^2}{m_{3/2}^2}\right) - 0.011 \left(1+3.062\,\frac{m_{1/2}^2}{m_{3/2}^2}\right) \ln\left(\frac{T_{\rm reh}}{10^{10}\,{\rm GeV}}\right)\right]\,.
 \label{yield1}
\end{align}
when we use $g_{\rm reh} = 915/4$ and $g(T\ll 1\,\text{MeV}) = 3.91$. Fig.~\ref{fig:Yapp} compares
the approximate result (\ref{yield1}) and the yield obtained from integrating (\ref{boltz2}) numerically 
with the full one-loop correction to the coupling constants $g_i$. We see that the agreement is excellent, 
within $\sim 10$\% for $T\gtrsim 10^{9}\,$GeV, after thermalization is completed.

The yield in (\ref{yield1}) can be rewritten in terms of $\Gamma_\phi$ once we specify
the `moment' of inflaton decay.  For example, If we assume that the decay occurs instantaneously at 
$\Gamma_\phi \, t = {2 \Gamma_\phi}/{3 H} = c$, where $c$ is a constant that is ${\cal O}(1)$, 
then we can write
\beq
T_{\rm reh} = \left( \frac{40}{g_{\rm reh} \pi^2} \right)^{1/4} \left( \frac{\Gamma_\phi M_P}{c} \right)^{1/2}\, .
\label{treh}
\eeq
We see that the dependence of (\ref{treh}) on the arbitrary parameter $c$ introduces an uncertainty in the gravitino abundance,
when expressed in terms of the physical decay rate of the inflaton. Inserting (\ref{treh}) into  (\ref{yield1}), and disregarding the logarithmic correction, one has 
\beq
Y_{3/2} \simeq \frac{0.00398}{\sqrt{c}} \, \left(\frac{\Gamma_\phi}{M_p}\right)^{1/2}  \, \left(1+0.558\,\frac{m_{1/2}^2}{m_{3/2}^2}\right) \, {\rm e}^{-\Gamma_{3/2} \, t} \, . 
\label{Y32-instant}
\eeq
As the inflaton does not decay instantaneously, there is in fact no `correct' value for $c$. Furthermore, 
the result (\ref{Y32-instant}) also assumes that all of the entropy produced by inflaton decays
is already present at $t = c/\Gamma_\phi$. As we will see below, in fact only about 1/3 of the entropy
produced by inflaton decays would have been released when $\Gamma_\phi t = 1$. Thus, an accurate determination
of the thermal gravitino yield after inflation requires the integration of the coupled
inflaton/radiation equations of motion. 

To summarize this Section:
using the thermal production rate computed in \cite{rs}, we have provided in Fig.~\ref{fig:Yapp} a numerical solution to gravitino yield as a function of time, 
as well as an analytical solution to the late-time yield. In the following Sections we study the extent to which gravitino production before the 
completion of thermalization modifies this standard calculation.


\section{Gravitino Production Assuming Instantaneous Thermalization of the Inflaton Decay Products}
\label{sec:precise-reh}

As discussed above, the previous calculation for the gravitino yield implicitly assumes the instantaneous 
decay of the inflaton at $\Gamma_\phi t \simeq 1$ and the
instantaneous thermalization of the inflaton decay products. In reality, the decay is a continuous process and
we now calculate the effect of gravitino production before reheating is complete, i.e., while the inflaton is still decaying, 
and still has a non-zero density $\rho_\phi$. We assume initially that the
decay products of the inflaton $\phi$ thermalize instantaneously, and discuss later the validity of this assumption.
With this assumption, the Universe contains, in addition to the undecayed inflaton fraction, a dilute thermal plasma whose  temperature is
\beq\label{inst_tempe}
T = \left(\frac{30\rho_{\gamma}}{\pi^2 g(T)}\right)^{1/4}\,,
\eeq
where $\rho_{\gamma}$ denotes the instantaneous energy density of the relativistic decay products. 
The time evolution of the full energy density during reheating is determined by the equations
\begin{align}\label{reh1}
\ddot{\phi} + (3H+\Gamma_{\phi})\dot{\phi} + V_{\phi}  &= 0\,,\\ \label{reh2}
\dot{\rho}_{\gamma} + 4H\rho_{\gamma} &= \Gamma_{\phi}\rho_{\phi}\,,\\ \label{reh3}
\rho_{\phi}+\rho_{\gamma} & = 3M_P^2H^2\,,
\end{align}
Reheating is complete when the energy density of the inflaton is negligible with respect to the density of the decay products,
\beq
\Omega_{\gamma} \equiv \frac{\rho_{\gamma}}{\rho_{\phi}+\rho_{\gamma}} = 1-\delta\,.
\label{delta}
\eeq
for some suitable $\delta\ll 1$. Since the total gravitino abundance is always relatively small,  
we can neglect the contribution of its decays to $\rho_{\gamma}$. Moreover, since the
oscillations of the inflaton about its minimum are much more rapid than any other time-scale in the problem, 
in particular since $m\gg \Gamma_{\phi}$,
Eq.~(\ref{reh1}) can be approximated by averaging the energy density of the inflaton over the oscillations. 
For the matter-like oscillations of the inflaton, 
$\langle \rho_{\phi}\rangle = \langle \dot{\phi}^2/2\rangle + \langle V\rangle \simeq \langle \dot{\phi}^2\rangle $, and therefore
(\ref{reh1}) reduces approximately to
\beq\label{reh4}
\dot{\rho}_{\phi} + 3H\rho_{\phi} = -\Gamma_{\phi}\rho_{\phi}\,.
\eeq
The solutions for $\rho_{\phi,\gamma}$ in terms of the scale factor are then given by
\begin{align}\label{rhophiex}
\rho_{\phi}(t) &= \rho_{\rm end}\left(\frac{a(t)}{a_{\rm end}}\right)^{-3}e^{-\Gamma_{\phi}(t-t_{\rm end})}\,,\\ \label{rhogex}
\rho_{\gamma}(t) & = \rho_{\rm end}\left(\frac{a(t)}{a_{\rm end}}\right)^{-4}\int_{\Gamma_{\phi}t_{\rm end}}^{\Gamma_{\phi}t} \left(\frac{a(t')}{a_{\rm end}}\right)e^{u_{\rm end}-u}\,du\,.
\end{align}
where the subscript indicates that the quantity is evaluated   at the end of inflation,
which is defined by the condition for the equation-of-state parameter $w \equiv p/\rho = -1/3$
or equivalently when $\dot \phi^2_{\rm end}/2 = V(\phi_{\rm end})$. We note that $\rho_{\rm end}$ is 
model-dependent and that $u_{\rm end} = \Gamma_\phi t_{\rm end}$. In the case of a Starobinsky-like potential \cite{Staro}
\beq
V(\phi) = \frac{3}{4}m^2\left(1-e^{-\sqrt{\frac{2}{3}}\phi} \right)^2
\label{staro-pot}
\eeq
we find $\rho_{\rm end}/m^2M_P^2=0.175$. 

\begin{figure}[h!]
\centering
    \psfrag{T}[b]{\LARGE{$T/(40\,\Gamma_{\phi}^2M_P^2/\pi^2g_{\rm reh})^{1/4}$}}
    \psfrag{v}{\LARGE{$v$}}
    \scalebox{0.6}{\includegraphics{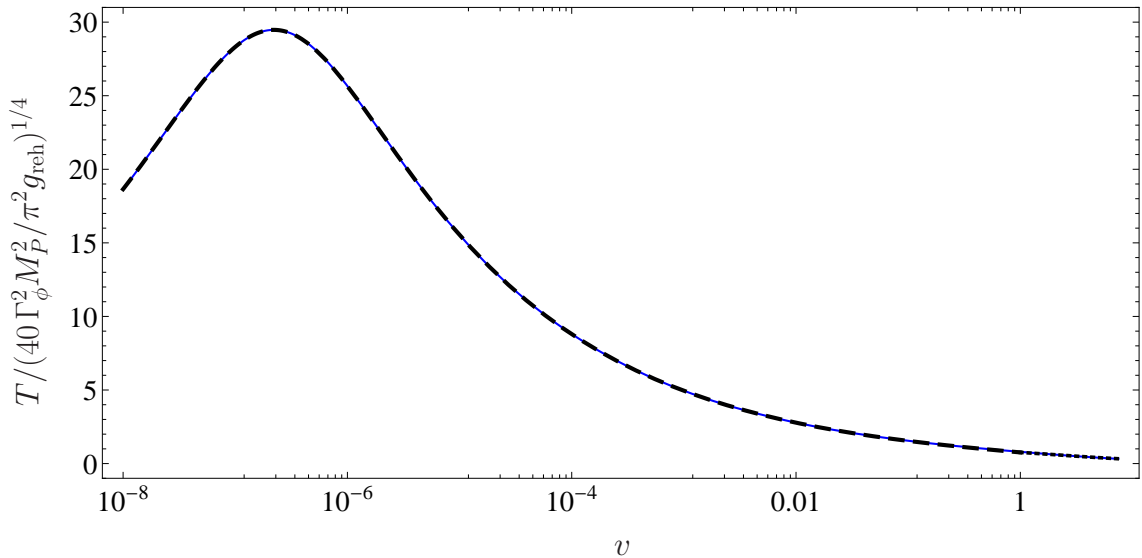}}
    \caption{\em The temperature of the dilute plasma that would be formed if   
    the inflaton decay products thermalized instantaneously, as a function of $v=\Gamma_{\phi}(t-t_{\rm end})$, for $\Gamma_{\phi}/M_P=10^{-12}$
 and $A \simeq 2.76\times10^{-7}$ (corresponding to the Starobinsky-like potential (\ref{staro-pot})). 
  The numerical solution of equations (\ref{reh1})-(\ref{reh3}) is shown as the solid blue curve. 
    The dashed (dotted) curve corresponds to the approximate solution (\ref{rhogapp2}) for $v<1$ ($v>1$). 
    All results are normalized relative to the reheating temperature derived from (\ref{rhoreh}). 
  We note the presence of the peak in the temperature at $v_{\rm max} \simeq 2.2\times10^{-7}$, see Eq. (\ref{vmax}). }
    \label{fig:maxT}
\end{figure}

The exact  solution of the Friedmann equation during reheating is given by
\beq\label{rhoexact}
\rho(t) = \rho_{\rm end}\left( 1 +  \sqrt{\frac{3}{4}\rho_{\rm end}}\,(1+\bar{w})\left(\frac{t-t_{\rm end}}{M_P}\right)\right)^{-2}\,
\eeq
or, equivalently, by
\beq
H(t) = \left(\frac{3}{2}(1+\bar{w})(t-t_{\rm end}) + H_{\rm end}^{-1}\right)^{-1}\,,
\eeq
where the time-averaged equation-of-state parameter $\bar{w}$ of the radiation/inflaton fluid is defined as
\beq
\bar{w}(t) \equiv \frac{1}{t-t_{\rm end}}\int_{t_{\rm end}}^{t}w(t')\,dt'\,.
\label{w-bar}
\eeq
For a slowly-varying $\bar{w}(t)$, the scale factor at times $t>t_{\rm end}$ can be approximated as
\beq\label{wgen}
\frac{a(t)}{a_{\rm end}} = \exp\left[\int_{t_{\rm end}}^t H(t')\,dt'\right] \simeq  \left( 1 +  \sqrt{\frac{3}{4}\rho_{\rm end}}(1+\bar{w})\left(\frac{t-t_{\rm end}}{M_P}\right)\right)^{\frac{2}{3(1+\bar{w})}}\,.
\eeq
During the initial stages of inflaton decay and thermalization, 
scalar field oscillations of the inflaton dominate and the equation-of-state parameter $w\approx0$. Substituting (\ref{wgen}) into (\ref{rhogex}), and introducing the quantities 
\beq
v\equiv \Gamma_{\phi}(t-t_{\rm end})\,, \qquad A\equiv \frac{\Gamma_{\phi}}{m}\left(\frac{3}{4}\frac{\rho_{\rm end}}{m^2M_P^2}\right)^{-1/2}\,,
\eeq
where $m$ denotes the inflaton mass, we can write the energy density of the relativistic decay products at early times as
\beq\label{rhogapp1}
\rho_{\gamma}  \simeq \rho_{\rm end}\left(\frac{v}{A}+1\right)^{-8/3}\int_{0}^{v} \left(\frac{v'}{A}+1\right)^{2/3}e^{-v'}\,dv'
\;\;\;\;,\;\; v \ll 1 \,.
\eeq
The solution (\ref{rhogapp1}) predicts a maximum of the energy density of the decay products: for $A\ll 1$ it corresponds to
\beq\label{vmax}
v_{\rm max} \simeq 0.80 A\qquad \Rightarrow \qquad \rho_{\gamma,{\rm max}} \simeq 0.21 A\rho_{\rm end}\,.
\eeq
This in turn implies a maximum temperature of the dilute plasma after the start of inflaton decay:
\beq\label{Tmax}
T_{\rm max} = 0.89 \left(\frac{A\rho_{\rm end}}{g(T_{\rm max})}\right)^{1/4}  \simeq 0.74 \left( \frac{\Gamma_\phi m M_P^2}{g_{\rm max}} \right)^{1/4} \,,
\eeq
where the second equality uses the value of $\rho_{\rm end}$ for the Starobinsky potential. 
Comparing $T_{\rm max}$ with the reheat temperature defined in (\ref{treh}),
we have
\beq
\frac{T_{\rm max}}{T_{\rm reh}} = 0.52 \left( \frac{g_{\rm reh}m}{g_{\rm max}\Gamma_\phi} \right)^{1/4}  \, ,
\label{tmaxreh}
\eeq
which correctly accounts for the peak of the temperature evolution shown in Fig.~\ref{fig:maxT}, when the inflaton mass is taken to be $m\simeq 10^{-5}M_P$.

During later stages of decay and reheating, the equation-of-state parameter of the radiation/inflaton fluid can be approximated by the average $\bar{w}(t_{\rm reh})$,
where $t_{\rm reh}$ is the time at which reheating completes. To be more precise, we denote by  $t_{\rm reh}$ the time at which the 
fractional energy density remaining in the inflaton field falls to a value $\delta \ll 1$, see Eq.~(\ref{delta}). 
The functional dependence of  $t_{\rm reh}$ on $\delta$, as well as the value of ${\bar w}$, can be obtained through the iterative 
procedure described in \cite{EGNO5}. Under the reasonable assumption that $A\ll1$,  one obtains 
$\Gamma_{\phi} \left( t_{\rm reh} - t_{\rm end} \right) \equiv v_{\rm reh} \simeq 0.655-1.082\ln\delta$, and $\bar{w}\simeq 0.273$ for $\delta = 0.002$ \cite{EGNO5}. 
One can then substitute (\ref{wgen}) with $w\simeq \bar{w}$ into  (\ref{rhogex}) to obtain $\rho_{\gamma}$ at late times. 
We can take into account the initial condition by matching the solution for $w=0$ (\ref{rhogapp1}) 
with the late solution for $w=\bar{w}$. Choosing the matching point at $v=1$ (corresponding to $t - t_{\rm end} =\Gamma_{\phi}^{-1}$), we obtain
\beq\label{rhogapp2}
\frac{\rho_{\gamma}}{\rho_{\rm end}}\simeq 
\begin{cases}
A^2e^A(v+A)^{-8/3}\left[\boldsymbol{\gamma}(\frac{5}{3},v+A)-\boldsymbol{\gamma}(\frac{5}{3},A)\right]\,, & v<1 \, ,\\[10pt]
1.44(1+\bar{w})^{-2}A^2v^{-8/3(1+\bar{w})}\boldsymbol{\gamma}(\frac{5+3\bar{w}}{3(1+\bar{w})},v)\,, & v>1 \, ,
\end{cases}
\eeq
where $\boldsymbol{\gamma}$ denotes the lower incomplete gamma function. The energy density at the end of reheating (more accurately, at $t = t_{\rm reh}$ defined above) may be evaluated from (\ref{rhoexact}), resulting in
\beq\label{rhoreh}
\rho_{\rm reh} \simeq \frac{4}{3} \left(\frac{M_P\Gamma_{\phi}}{(1+\bar{w}) v_{\rm reh}} \right)^2 \,.
\eeq
Fig.~\ref{fig:maxT} shows the temperature evolution during reheating obtained from the approximate solution (\ref{rhogapp2})
and from the numerical solution of equations (\ref{reh1})-(\ref{reh3}), where the scalar potential has been chosen as the 
Starobinsky potential: we see excellent agreement. 

\begin{figure}[h!]
\centering
    \psfrag{S}[b]{\LARGE{$S/S_{\rm final}$}}
    \psfrag{v}{\LARGE{$v$}}
    \scalebox{0.6}{\includegraphics{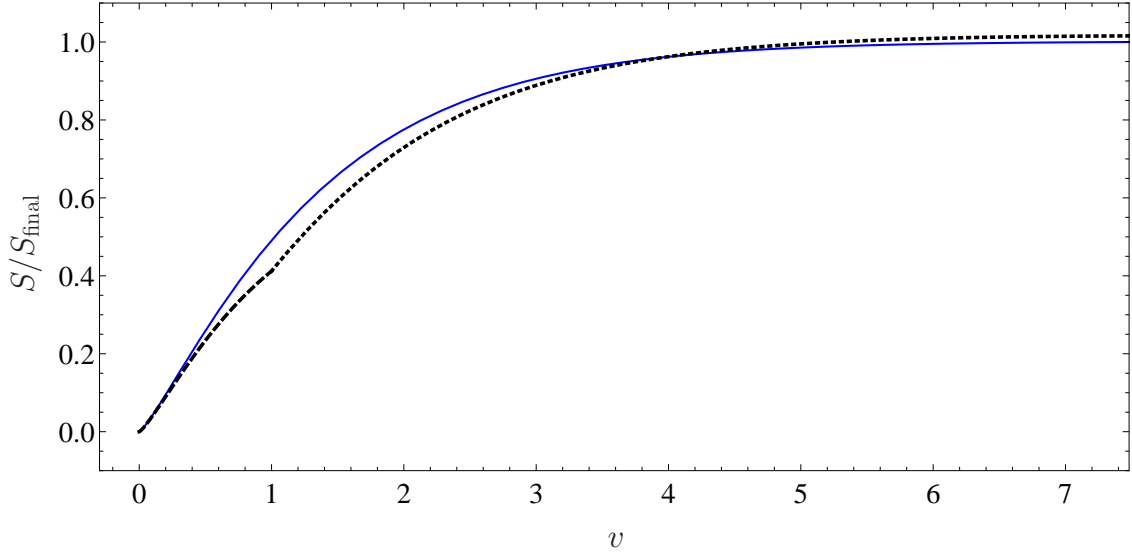}}
    \caption{\em The entropy $S=sa^3$ as a function of $v=\Gamma_{\phi}(t-t_{\rm end})$, for the same values of $
   \Gamma_\phi$ and $A$ as in the previous figure. We note that
    reheating is still incomplete when $v=1$, and that the net entropy is conserved after reheating ends, 
    for $\delta\ll 1$. The solid blue line is the numerical result, and the dashed (dotted) curve corresponds to the 
    approximate solution (\ref{rhogapp2}) for $v<1$ ($v>1$).}
    \label{fig:entropy}
\end{figure}

For a brief period of time, the temperature is significantly larger
than $T_{\rm reh}$ and, as seen in (\ref{tmaxreh}), the relative increase in $T$ scales with
$m/\Gamma_\phi$. Since the production rate of gravitinos is proportional to $T$, 
one might suspect that gravitino production at $v < 1$ could contribute substantially to the total gravitino abundance.
However, one must also take into account the growth of the entropy $S=sa^3$ during the epoch of inflaton decay and reheating illustrated in Fig.~\ref{fig:entropy}, where we show the entropy normalized to its final value $S_{\rm final}$. We note that the entropy is relatively small during the period of the bump in temperature seen in Fig.~\ref{fig:maxT}.
Although the plasma is hot at this time, any production at $v < 1$ will be diluted subsequently by later inflaton decays that builds up entropy, as seen in Fig.~\ref{fig:entropy}.  
Thus, the net effect of the increased temperature and later dilution is not obvious {\it a priori}.

Gravitinos are produced continuously by collisions of the relativistic decay products throughout this reheating epoch. 
However, the gravitino yield cannot be computed directly from (\ref{boltz3}), 
due to the continuing injection of entropy into the plasma while the decay of the inflaton continues, 
as illustrated in Fig \ref{fig:entropy}. We therefore make use of the full equation (\ref{boltz2}), 
which can be rewritten in terms of $v$ as 
\beq\label{ysc}
Y_{3/2}' + 3\left(\hat{H}+\frac{\hat{T}'}{\hat{T}}\right)Y_{3/2} = \left(\frac{\Gamma_{\phi}}{M_P}\right)^{1/2}\hat{\Sigma}\,\hat{n}_{\rm rad} \, ,
\eeq
where $\hat{H} \equiv H/\Gamma_{\phi}$, $\hat{T} \equiv T/(\Gamma_{\phi}M_P)^{1/2}$, $\hat{n}_{\rm rad}
\equiv n_{\rm rad}/(\Gamma_{\phi}M_P)^{3/2}$ and $\hat{\Sigma} \equiv \langle\sigma_{\rm tot}v_{\rm rel}\rangle M_P^2$. Note that for $v\gg A$, we can approximate Eq.~(\ref{rhoexact}) by
\beq
\rho(v) \simeq \frac{4}{3} \left(\frac{M_P\Gamma_{\phi}}{(1+\bar{w}) v} \right)^2 \equiv (\Gamma_{\phi}M_P)^2\hat{\rho}(v) \,.
\eeq
For a constant number of degrees of freedom during reheating, these rescalings  allow us to rewrite Eq.~(\ref{ysc}) as
\beq
\begin{aligned}\label{ysc2}
Y_{3/2}' + 3\left[ \left(\frac{\hat{\rho}_{\phi}+\hat{\rho}_{\gamma}}{3}\right)^{1/2} + \frac{\hat{\rho}'_{\gamma}}{4\hat{\rho}_{\gamma}} \right]Y_{3/2} = \;&\frac{3}{16\pi}\left(\frac{30\hat{\rho}_{\gamma}}{\pi^2}\right)^{3/4}\\  
&\times g^{-3/4}\left(\frac{\Gamma_{\phi}}{M_P}\right)^{1/2} \sum_{i=1}^3 c_i\, g_i^2\left(1+\frac{m_{\tilde{g}_i}^2}{3m_{3/2}^2}\right)\ln\left(\frac{k_i}{g_i}\right).
\end{aligned}
\eeq
Disregarding the temperature dependence of the second line, (\ref{ysc2}) has the formal solution 
\beq
\begin{aligned}\label{yasol0}
Y_{3/2}(v) =\;& \frac{3}{16\pi}\left(\frac{30}{\pi^2}\right)^{3/4}\int_0^{v} \hat{\rho}_{\gamma}(u)^{3/4}\exp\left[-3\int_{u}^{v}\left\{ \left(\frac{\hat{\rho}(z)}{3}\right)^{1/2} + \frac{\hat{\rho}'_{\gamma}(z)}{4\hat{\rho}_{\gamma}(z)} \right\}\,dz\right]\,du\\
&\qquad\qquad\qquad\qquad\qquad \times g^{-3/4}\left(\frac{\Gamma_{\phi}}{M_P}\right)^{1/2} \sum_{i=1}^3 c_i\, g_i^2\left(1+\frac{m_{\tilde{g}_i}^2}{3m_{3/2}^2}\right)\ln\left(\frac{k_i}{g_i}\right).
\end{aligned}
\eeq
The integrand of this equation provides the contribution to the abundance at the (rescaled) time $v$ of the gravitinos produced at any moment $u$ between $0$ and $v$. For any value of $u$, the exponential factor accounts for the dilution of those gravitinos due to the inflaton decay, from the times they are produced ($u$) to the time at which the abundance is evaluated ($v$). 
The $v$-dependence for $v\gg1$ may be extracted by splitting the integral in the radiation-dominated era, i.e. integrating over $u$ from $0$ to $\infty$, and then subtracting the contribution from $v\gg 1$ to $\infty$. Since the argument of the exponential dilution factor vanishes during radiation domination, the result can be written as
\begin{align}
\int_0^{v} &\hat{\rho}_{\gamma}(u)^{3/4}\exp\left[-3\int_{u}^{v}\left\{ \left(\frac{\hat{\rho}(z)}{3}\right)^{1/2} + \frac{\hat{\rho}'_{\gamma}(z)}{4\hat{\rho}_{\gamma}(z)} \right\}\,dz\right]\,du  \notag\\ 
&=\int_0^{\infty} \hat{\rho}_{\gamma}(u)^{3/4}\exp\left[-3\int_{u}^{\infty}\left\{ \left(\frac{\hat{\rho}(z)}{3}\right)^{1/2} + \frac{\hat{\rho}'_{\gamma}(z)}{4\hat{\rho}_{\gamma}(z)} \right\}\,dz\right]\,du\,\ -\, \int_v^{\infty} \hat{\rho}_{\gamma}(u)^{3/4}\,du \notag \\ \label{split}
&=\int_0^{\infty} \hat{\rho}_{\gamma}(u)^{3/4}\exp\left[-3\int_{u}^{\infty}\left\{ \left(\frac{\hat{\rho}(z)}{3}\right)^{1/2} + \frac{\hat{\rho}'_{\gamma}(z)}{4\hat{\rho}_{\gamma}(z)} \right\}\,dz\right]\,du\,\ -\, 2\left(\frac{3}{4}\right)^{3/4}v^{-1/2}\,.
\end{align}
During reheating, the energy densities may be approximated by (\ref{rhoexact}) and (\ref{rhogapp2}). An improved estimate, valid for any $A< v<\infty$, can be constructed if one considers (\ref{rhogapp2}) a zeroth-order approximation to $\rho_{\gamma}$. The first-order estimates of the energy densities can then be built by approximating the time-dependent average equation-of-state parameter by  
\beq
\bar{w}(v)\simeq 
\frac{1}{v} \int_0^v d u \frac{\rho_\gamma \left( u \right) / 3}{\rho_\gamma \left( u \right) + \rho_\phi \left( u \right)} \Big\vert_{w=0} = 
 \frac{1}{3v}\int_{0}^v \frac{\boldsymbol{\gamma}(\frac{5}{3},u)}{\boldsymbol{\gamma}(\frac{5}{3},u) + u^{2/3}e^{-u}}\,du\,,
\eeq
and substituting it into (\ref{rhophiex}), (\ref{rhoexact}) and (\ref{wgen}). With this procedure, the first term of (\ref{split}) is evaluated to be 1.7, which implies that
the solution of equation (\ref{ysc2}) may be 
approximated for $A\ll 1$ by 
\beq\label{yc1}
Y_{3/2} \left( T \right) \;\simeq\; (0.233-0.221\,v^{-1/2})\, g^{-3/4}\left(\frac{\Gamma_{\phi}}{M_P}\right)^{1/2} \sum_{i=1}^3 c_i\, g_i(T_{\rm reh})^2\left(1+\frac{m_{\tilde{g}_i}^2}{3m_{3/2}^2}\right)\ln\left(\frac{k_i}{g_i(T_{\rm reh})}\right)
\eeq
after the end of reheating, while the temperature of the thermal bath is still high enough that the number of relativistic degrees of freedom $g=915/4$.

\begin{figure}[h!]
\centering
    \psfrag{C}[b]{\Large{$Y_{3/2}/\left(\frac{\Gamma_{\phi}}{M_P}\right)^{1/2}$}}
    \psfrag{G}{\Large{$\Gamma_{\phi}/M_P$}}
    \scalebox{0.62}{\includegraphics{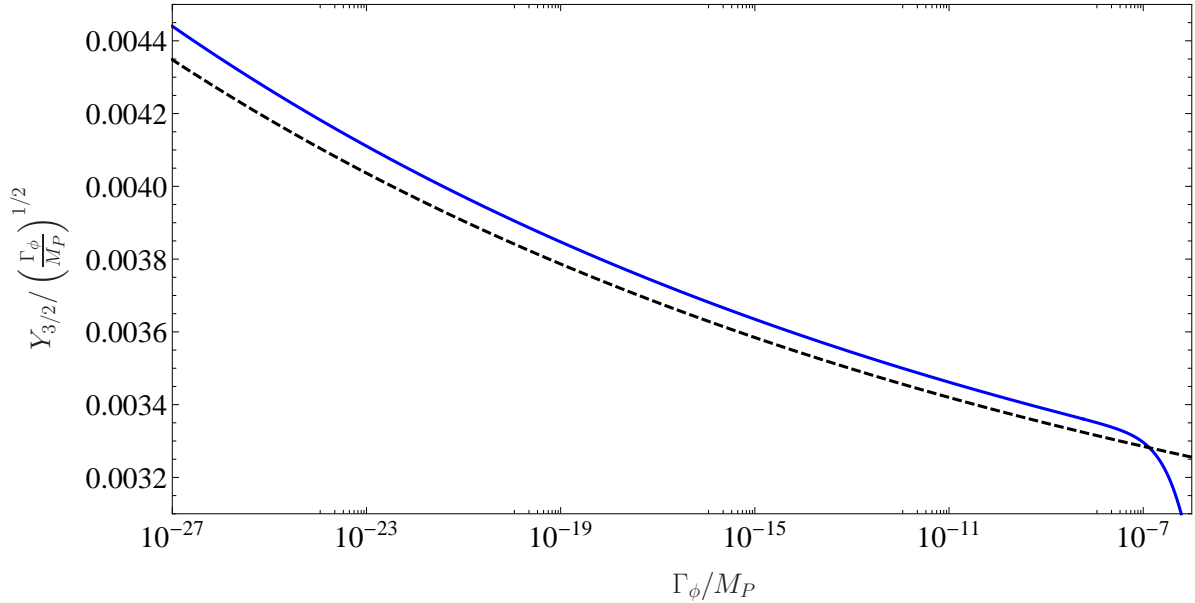}}
    \caption{\em The final gravitino yield $Y_{3/2}(T\ll 1\,{\rm MeV})$ as a function of the inflaton decay rate, assuming instantaneous 
    thermalization during inflaton decay, for the $m_{1/2}$-independent transversal components. The normalization is chosen to emphasize the dependence of $Y_{3/2}$ 
    on the running of the coupling constants. The blue continuous line corresponds to the numerical solution obtained from
    equations (\ref{reh1})-(\ref{reh3})
    . The black dashed line shows the approximation (\ref{yc1}).  Here we assume the MSSM value $g=915/4$ for the number of degrees of freedom during reheating.  
}
    \label{fig:cy}
\end{figure}
%

Figure \ref{fig:cy} displays the comparison between (\ref{yc1}) and the exact numerical result,
showing that they are in agreement within $\lesssim 2\%$ over the entire range $\Gamma_\phi  \lesssim  10^{-7} \, M_p$. 
The final abundance for $T\ll 1\,\text{MeV}$ is obtained in the limit  $v \gg 1$ in (\ref{yc1}), including the dilution factor $g(T)/g_{\rm reh}$, 
%
\begin{align}\label{yc2}
Y_{3/2} (T ) \;&\simeq\; 0.233\;\frac{g(T)}{g_{\rm reh}^{7/4}}\, \left(\frac{\Gamma_{\phi}}{M_P}\right)^{1/2} \sum_{i=1}^3 c_i\, g_i(T_{\rm reh})^2\left(1+\frac{m_{\tilde{g}_i}^2}{3m_{3/2}^2}\right)\ln\left(\frac{k_i}{g_i(T_{\rm reh})}\right) \\ 
&\simeq\; 0.00363 \left(1+0.56\,\frac{m_{1/2}^2}{m_{3/2}^2}\right) \, \left(\frac{\Gamma_\phi}{M_p}\right)^{1/2}  \,, \label{ytreh2}
\end{align}
where we have assumed in the second line that $g_{\rm reh} = 915/4$, $g(T\ll 1\,\text{MeV}) = 3.91$, and $T_{\rm reh}\sim10^{10}$\,GeV for the couplings, neglecting logarithmic corrections. 
This result can be compared to the full numerical solution of (\ref{ysc}), which takes into account the running of the gauge couplings, integrated from the beginning of reheating to $v\gg v_{\rm reh}$, 
deep into the radiation-dominated era, where the gravitino yield asymptotes to its final value. Accounting for the dilution factor, a phenomenological fit to this numerical result gives 
\beq\label{y_final}
Y_{3/2} (T ) \;=\; 0.00360 \, \left(\frac{\Gamma_{\phi}}{M_P}\right)^{1/2}
\eeq
for the $m_{1/2}$-independent part in the range $10^{-20}\leq \Gamma_{\phi}/M_P\leq 10^{-8}$. 

\begin{figure}[h!]
\centering
    \psfrag{Y}[b]{\LARGE{$Y_{3/2}/\left(\frac{\Gamma_{\phi}}{M_P}\right)^{1/2}$}}
    \psfrag{v}{\LARGE{$v$}}
    \scalebox{0.6}{\includegraphics{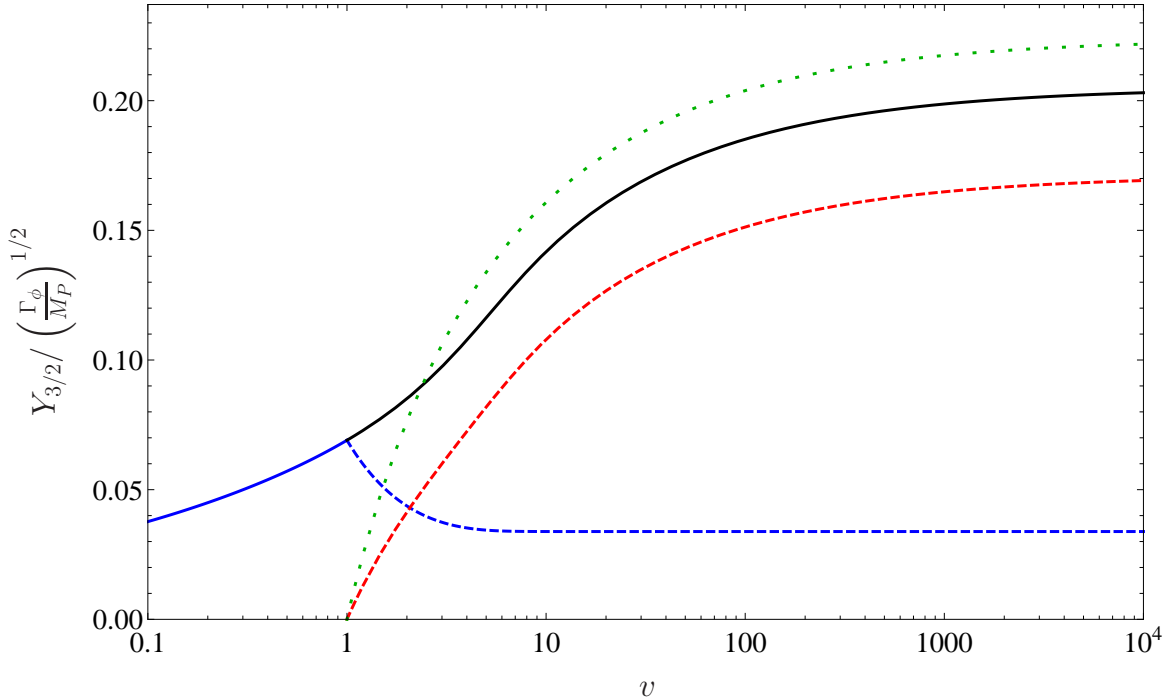}}
    \caption{\em The evolution of the gravitino abundance, as a function of $v=\Gamma_{\phi}(t-t_{\rm end})$ during and after reheating, for the decay rate $\Gamma_{\phi} = 10^{-12}M_P$. The solid curve corresponds to the numerical solution of (\ref{ysc}), assuming $g_{\rm reh}=915/4=$const. (i.e. no dilution due to the factor $g(T)/g_{\rm reh}$ has been included), and neglecting $m_{1/2}$-dependent terms in the collision term. The blue curve shows the evolution of the yield produced only at $v<1$; the dilution of that yield for $v>1$ is shown in the blue dashed curve. The red dashed curve tracks the abundance produced at $v>1$, and the sum of the dashed curves is shown as the black solid curve. The dotted green curve demonstrates the evolution of the gravitino yield assuming instantaneous decay and thermalization at $v=1$.  
}
    \label{fig:yfull}
\end{figure}

We are now in a position to compare our exact result (\ref{y_final}) with the naive approximation 
made previously in Section \ref{sec2}.  In Fig. \ref{fig:yfull} we show the evolution of the gravitino 
abundance as a function of time (as parametrized by $v$). The solid curve shows the exact solution,
which asymptotes to $Y_{3/2}/\left(\Gamma_{\phi}/M_P\right)^{1/2}\simeq 0.203$. Since $T(v=10^4)\sim 10^{10}\,{\rm GeV}$ for the decay rate considered, this result would still need to be multiplied by the dilution factor $g(T)/g_{\rm reh}\simeq 0.0171$ in order to obtain the final abundance, $Y_{3/2}(T\ll T_{\rm reh})\simeq 3.44\times 10^{-9}$, in very good agreement with (\ref{ytreh2}) and (\ref{y_final}), which give $Y_{3/2}\simeq 3.63\times 10^{-9}$ and $3.60\times 10^{-9}$, respectively (in comparing with these relations, recall that  $\Gamma_{\phi} = 10^{-12}M_P$ is assumed in the example shown in the Figure).
The blue curve tracks the abundance at $v<1$, and the dilution of that production is tracked by the 
dashed blue curve at $v>1$. Production (and dilution) at $v>1$ is shown by the dashed red curve,
and the sum of the two dashed curves gives the solid black curve. 
The result (\ref{y_final}) demonstrates that the gravitino abundance is sensitive primarily to the final reheating temperature, after the production of entropy has ceased, 
rather than to the maximum temperature of the Universe seen in Fig.~\ref{fig:maxT}, as originally pointed out in \cite{Giudice:1999am}. 

\begin{figure}[h!]
\centering
    \psfrag{R}[b]{\LARGE{$Y_{3/2}^{\rm naive}/Y_{3/2}^{\rm exact}$}}
    \psfrag{v}{\LARGE{$c$}}
    \scalebox{0.6}{\includegraphics{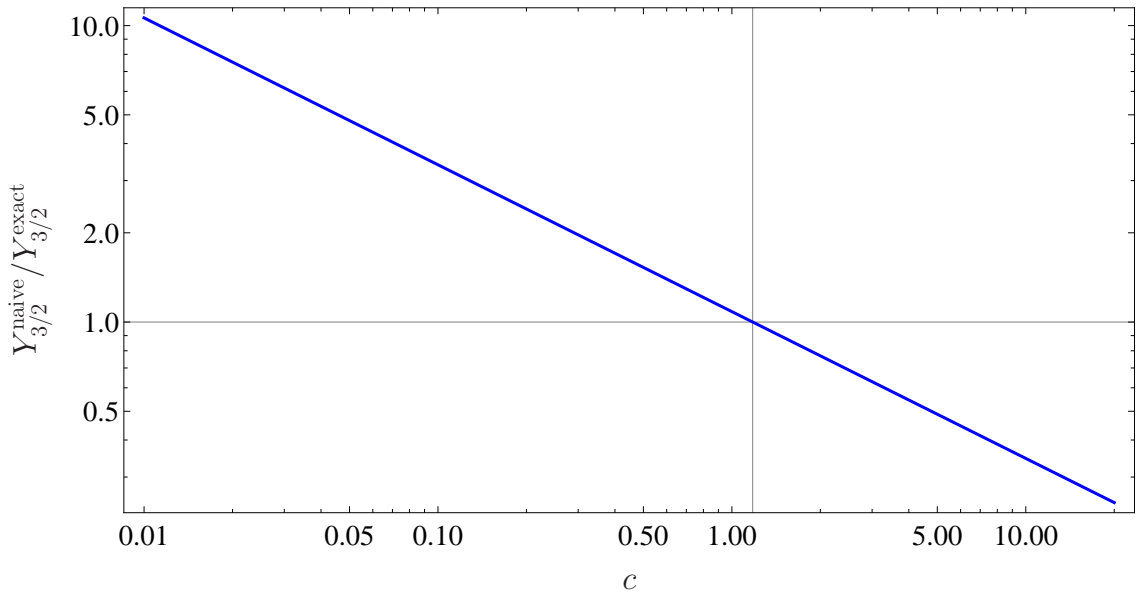}}
    \caption{\em The ratio of the final gravitino yield $Y_{3/2}^{\rm naive}$, assuming instantaneous decay and thermalization at $v=c$, to the exact yield. The results agree for $c\simeq 1.2$.}
    \label{fig:ratio}
\end{figure}

The exact result should be compared
with the green dotted curve that assumes instantaneous decay at $v = 1$,  namely $c=1$ in eqs. (\ref{treh}) and (\ref{Y32-instant}).
Impressively, at large $v$ these results lie within 10\% of each other~\footnote{The difference between the asymptotic value for the yield of $0.222$ for $c=1$ shown in Fig.~\ref{fig:yfull} and the expected value of $0.233$ from (\ref{Y32-instant}) is due to the logarithmic correction in (\ref{yield0}).}. As noted earlier, another common choice for the instantaneous decay is $c = 2/3$ (namely $\Gamma_\phi = H$), and in this case we would have an asymptotic yield of 0.272,  which would correspond to $Y_{3/2}(T\ll T_{\rm reh})\simeq 4.65 \times 10^{-9}$. The ratio of the naive  (\ref{Y32-instant})  to the exact (\ref{y_final}) result is shown in Fig. \ref{fig:ratio} as a function of $c = \Gamma_\phi \, t_{\rm inst. \, reh.}$. We see that the instantaneous approximation yields the correct result when  $c \simeq 1.2$ \cite{ps2}.


So far we have assumed that thermalization occurs instantaneously upon the end of inflation. However, since the initial distribution of the decay products of the inflaton is not thermal, the thermalization of the relativistic plasma will not be completed until the interactions of the constituent particles are sufficient to create a thermal distribution. In the next Section we study in more detail the approach to kinetic and chemical equilibrium, finding that, for a small decay rate $\Gamma_{\phi}\ll m$, number-conserving and number-changing processes at small scattering angles would efficiently thermalize the plasma well before the end of reheating, $v_{\rm th}\ll v_{\rm reh}$. It is in any case important to quantify the dependence of the final gravitino abundance on the thermalization rate.  The comparison between (\ref{Y32-instant}), with $c = {\rm O } \left( 1 \right)$, and  (\ref{y_final}) suggests that the bulk of the relic gravitino density is produced at $v\sim 1$, and therefore $Y_{3/2}$ may only have a weak dependence on $v_{\rm th}$. Nonetheless, a non-negligible fraction of the total gravitino abundance may still be produced at $v\ll 1$, since at early times $T>T_{\rm reh}$.

\begin{figure}[h!]
\centering
    \psfrag{Y}[b]{\LARGE{$Y_{3/2}^{\rm delayed}/Y_{3/2}^{\rm inst}$}}
    \psfrag{v}{\LARGE{$v_{\rm th}$}}
    \scalebox{0.6}{\includegraphics{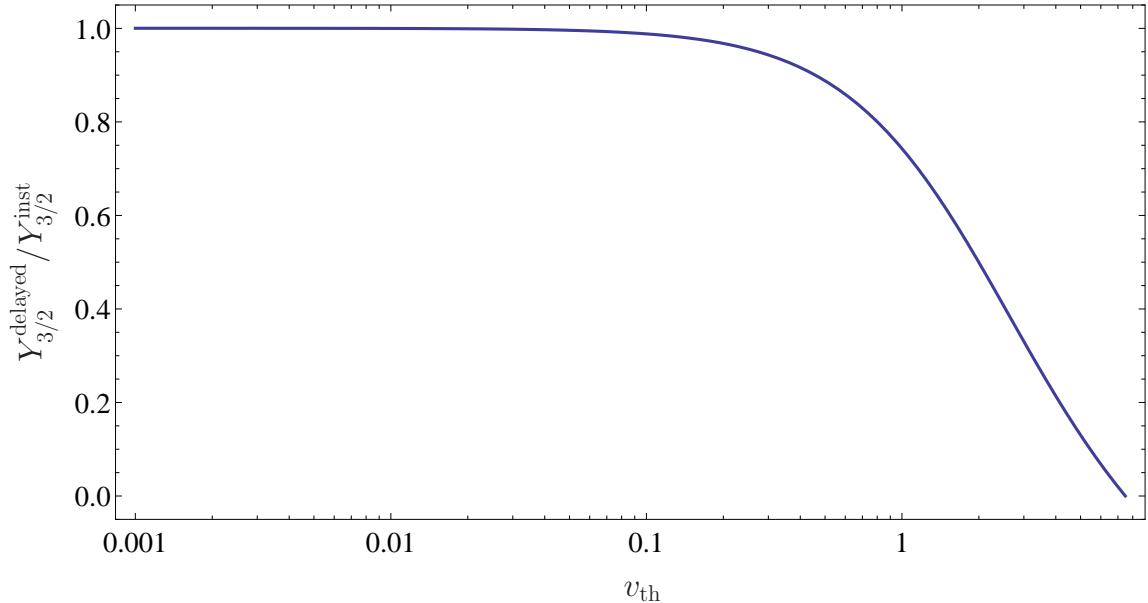}}
    \caption{\em The ratio of the gravitino yield at the end of reheating $Y_{3/2}(T_{\rm reh})$ assuming delayed thermalization at $v_{\rm th}$, to the yield assuming instantaneous thermalization, as a function of the thermalization delay $v_{\rm th}$. For $Y_{3/2}^{\rm delayed}$, production at $v<v_{\rm th}$ is ignored. The suppression of the final abundance is negligible unless $v_{\rm th}\gtrsim 0.1$}
    \label{fig:thdl}
\end{figure}

We assume here that the distribution of the inflaton decay products is non-thermal before $v_{\rm th}$, and we disregard gravitino
production at this early stage, thus obtaining a lower bound on the total amount of produced gravitinos. 
Without entering in the details of thermalization (which are discussed in the next Section), here we treat $v_{\rm th}$ as a free parameter.
The goal of this analysis is to understand how much the final gravitino abundance is dependent on this parameter and, ultimately, 
how much our final result (\ref{y_final}) is affected by possible uncertainties on the thermalization processes.  The results summarized in
Fig.~\ref{fig:thdl} show that the final abundance  (\ref{y_final})  is extremely robust. This result holds with very good accuracy provided that  
$v_{\rm th} \lesssim 0.1$. Essentially, only the gravitinos produced when $v \gtrsim 0.1$ contribute to the final abundance,
as the gravitinos produced at earlier stages are diluted away. It is possible that thermalization has not taken place by the time 
$v_{\rm max} \ll 1$ given in Eq.~(\ref{vmax}), so that it is possible that the maximum temperature (\ref{Tmax}), is not reached 
(we discuss this in detail in Subsection \ref{subsec:smallangle}). However, this does not affect the validity of   (\ref{y_final}). 
For the result  (\ref{y_final}) to be valid, we only need to assume that the inflaton decay products are thermalized by $v \simeq 0.1$. 
In the next Section we verify that this is indeed the case. 

%

\section{The Thermalization Process}
\label{sec:therma}

In the previous sections we have assumed that the inflaton decay products thermalize instantaneously.
We now discuss this approximation, studying the relevant physical processes and the timescale for the thermalization. In the limit in which the created particles are not interacting, their number density will be given by
\beq\label{nit}
n_p(t) \simeq \frac{\rho_{\rend}}{m}\left(1-e^{-\Gamma_{\phi}(t-t_{\rm end})}\right)\times\left(\frac{a(t)}{a_{\rend}}\right)^{-3}\,,
\eeq
up to an order one factor that we disregard here. This number density should be compared to the thermal one, given by 
\beq
n_{\rm th}(t) = \frac{\zeta(3)}{\pi^2}\tilde{g}(T)\left(\frac{30\rho_{\gamma}(t)}{\pi^2 g(T)}\right)^{3/4} \, ,
\eeq
where $\tilde{g} \equiv \sum_B g_B + \frac{3}{4}\sum_F g_F$ denotes the effective number of relativistic degrees of freedom
contributing to the number density.  From this result and from (\ref{rhogapp2}) we obtain 
\begin{eqnarray} 
\frac{n_{\rm th}}{n_p} &\;\simeq\;& 0.28 \, \frac{\tilde{g}(T)}{g(T)^{3/4}}\frac{m}{\rho_{\rm end}^{1/4}} \, \frac{
{\rm e}^{\frac{3 A}{4}} \, 
\left(\boldsymbol{\gamma} \left( \frac{5}{3} , A+v \right) - \boldsymbol{\gamma} \left(\frac{5}{3} , A \right)  \right)^{3/4}}{\sqrt{A} \left( 1 - {\rm e}^{-v} \right)}  
 \nonumber\\ 
&\;\simeq\;& 0.26 \frac{\tilde{g}(T)}{g(T)^{3/4}}\frac{m}{(\Gamma_{\phi}M_P)^{1/2}}\frac{
\boldsymbol{\gamma} \left( \frac{5}{3} , v \right)^{3/4}}{ 1 - {\rm e}^{-v} } \,, 
\label{ratio}
\end{eqnarray} 
where the first expression assumes $v \lesssim 1$, while the second expression assumes     $A \ll v \lesssim 1$. 

We can simplify this expression by noting that the last factor in (\ref{ratio}) interpolates between $0.68 \, v^{1/4}$ at $v \ll 1$ and $0.69$ at $v=1$. We then use  $g(T) = 915/4$, $\tilde{g}(T) = 427/2$, and we parametrize  the inflaton decay rate  as $\Gamma_{\phi}=m|y|^2/8\pi$. We obtain 
\beq 
\frac{n_{\rm th}}{n_p} \;\simeq\; 3.2 \, v^{1/4} \, \frac{\sqrt{m}}{\vert y \vert \, \sqrt{M_P}} \;\;\;,\;\;\;   A \ll v \lesssim 1 \,. 
\eeq 
Assuming $m \simeq 10^{-5}M_P$, 
we thus see that, if they do not interact,  the decay products are in a regime of `under-occupation' with respect to the thermal case (namely, 
 $n_{\rm th}/n_p>1$) for $|y| \lesssim {\mathcal O } \left( 10^{-2} \right) v^{1/4}$; since $v<v_{\max}\sim |y|^2/8\pi$, this is equivalent to $|y| \lesssim {\mathcal O } \left( 10^{-5} \right)$ for $v>v_{\rm max}$. This is typically the case, for example, in no-scale models~\cite{EGNO5}, when the bulk of the inflaton quanta decay.   We conclude that processes that increase the number of quanta need to be effective for thermalization to take place. We now consider some mechanisms for thermalization of the inflaton decay products.

\subsection{Large-Angle Scattering}

As we discussed in the previous section, number-increasing processes are necessary to thermalize the relativistic decay products of the inflaton. However, elastic scatterings are also of interest, since they can bring the plasma into kinetic equilibrium. 

We start by considering the $2 \leftrightarrow 2$ elastic scatterings among the decay products. The cross section decreases as the square of the transfer momentum, which, at large angle, is of the order of the momentum of the incoming particles, $\sigma\sim \alpha^2/p^2$, where $\alpha$ denotes the coupling strength. Therefore the rate is greatest  for the particles produced at the earliest time, once their energy redshifts from the initial ${\mathcal O } \left( m \right)$ energy to
\begin{equation}
\label{prange}
p_{\rm redshifted} \simeq \frac{m}{2} \, \frac{a_{\rm end}}{a \left( t \right)} \simeq \frac{m^{1/3}}{2 \, \left( t - t_{\rm end} \right)^{2/3}}\,, \qquad  t - t_{\rm end}\gg m^{-1} \,. 
\end{equation} 
One could then imagine that the redshifted inflaton decay products can thermalize among themselves, and that more energetic particles, produced by later inflaton decay, can  in turn thermalize by interacting with this soft, thermalized tail 
after it has reached equilibrium~\cite{McDonald:1999hd}.

However, one needs to take into account that, at any moment during the decay of the inflaton,  hard particles with $p\sim m$ 
(namely, those produced around that moment) are more abundant than the previously-produced redshifted decay products. At any given time, the spectrum of the decay products can be found 
approximately by solving the Boltzmann transport equation in the absence of interactions:
\beq\label{boltz0}
\frac{\partial f_p}{\partial t} - Hp\frac{\partial f_p}{\partial p} \simeq 0\,,
\eeq
where $f_{p}(t)$ denotes the distribution density of the decay products, $n_p = \frac{g_p}{(2\pi)^3}\int d^3p\,f_p$. When thermalization is complete this distribution density depends on time only through the temperature, $f_p\sim e^{-p/T}$. The solution of equation (\ref{boltz0}) that matches the early time limit of (\ref{nit}) using (\ref{wgen}), $A\ll v\ll 1$, namely 
\begin{equation}
n_p \simeq \frac{4 \, \Gamma_\phi^2 \, M_p^2}{3 \, m \, v}\,,
\label{ni-early}
\end{equation}
is 
\beq\label{f_hard}
f_p(t) = \frac{8\sqrt{2}\pi^2}{g_p}\,  \frac{\Gamma_{\phi}^2 M_P^2}{m^4} \, 
\left(\frac{m}{p}\right)^{3/2} \, v^{-1} \,,
\eeq
which implies a spectrum of the form 
$\frac{d n}{d p} \sim \frac{\Gamma_\phi^2 M_p^2 p^{1/2}}{m^{5/2} v}$. It follows that the large-angle scattering cross section is IR dominated: 
\beq
\frac{\Gamma_{\rm elas}}{H} \;=\; \frac{1}{H} \int \sigma \,dn  \;\simeq\; \frac{1}{H} \int_{p_{\rm redshifted}}^{\frac{m}{2}} dp\,\frac{dn}{dp}\frac{\alpha^2}{p^2} \;\simeq \;
 \alpha^2\left(\frac{M_P}{m}\right)^2\left(\frac{\Gamma_{\phi}}{m}\right)^{2/3}v^{1/3}\,. 
\label{gHn}
\eeq
Assuming $m\simeq 10^{-5}M_P$ and scattering mediated by the strong interaction, the ratio (\ref{gHn}) shows that $2\leftrightarrow2$ processes can achieve kinetic
equilibrium at $v<1$, if the decay rate is $\Gamma_{\phi}\gtrsim 0.05\,m^3/M_P^2$. However, in some no-scale models the inflaton decay rate lies below this bound \cite{EGNO4}.

When the rate of the $2\leftrightarrow2$ scatterings is too small,  processes that increase particle number
must be effective, not only to increase the overall number density to the thermal one, but also to produce soft particles efficiently,
if one hopes to thermalize the plasma before the end of reheating. At large angles, the rate for $2\rightarrow 3$ splitting, 
which is the most efficient inelastic process, may be approximated by $\sigma\sim \alpha^3/p^2$~\cite{Davidson:2000er}.
We can then mirror the analysis leading to (\ref{gHn}), but with a rate further suppressed by an extra factor of $\alpha$. Under the same assumptions, thermalization would occur before $v=1$ only if $\Gamma_{\phi}\gtrsim 4.4\,m^3/M_P^2$.
We conclude that, for a decay rate of the inflaton arising from Planck-suppressed interactions,
thermalization before the end of reheating generically does not occur via large-angle scattering processes, since (1) not enough particles can be created to populate the thermal bath, and (2) the spectrum of the existing decay products remains too hard. However, as we discuss in the following subsection, these two shortcomings can be cured by considering the small-angle scattering of the inflaton decay products.

\subsection{Small-Angle Scattering}
\label{subsec:smallangle}

At small angles, the momentum transfer vanishes, and so the cross section is in general infrared divergent in the vacuum. 
Due to the increase of the rate, a more careful kinetic-theory approach than the one used in the previous subsection 
has to be adopted \cite{Kurkela:2011ti}. The Boltzmann equation controls the 
evolution of the (uncorrelated) one-particle distribution functions under the assumption that the relaxation time is much larger 
than the collision time, so there is no interference between successive scatterings. However,
this assumption is not valid in gauge theories, as the mean free time between small-angle scatterings in a thermal bath
can be shown to be of the same order as the time for formation of a bremsstrahlung gauge boson, 
$\tau\sim 1/g^2T$~\cite{Arnold:2007pg}. This results in a suppression of the bremsstrahlung rate, 
since now several successive small-angle collisions become virtually indistinguishable from a single collision
(the Landau-Pomeranchuk-Migdal (LPM) effect)~\cite{LPM}. 
Nevertheless, the introduction of the full
Bogoliubov-Born-Green-Kirkwood-Yvon (BBGKY) hierarchy of equations for a system of interacting particles
can be avoided by setting up an effective Boltzmann equation, which can be written schematically as \cite{Kurkela:2011ti}

\begin{align}
\notag
\frac{\partial f_p}{\partial t} - Hp\frac{\partial f_p}{\partial p}\; &=\; \left|
   \begin{picture}(60,20) (0,0)
    \SetWidth{1.0}
    \SetColor{Black}
    \Line[](2,17)(56,17)
    \Line[](2,-10)(56,-10)
    \Gluon(29,17)(29,-10){2.0}{5}
  \end{picture}
  \right|^2\; +\; \left|
  \begin{picture}(60,20) (0,0)
    \SetWidth{1.0}
    \SetColor{Black}
    \Line[](2,-2)(56,-2)
    \Gluon(29,8)(56,8){2.0}{5}
    \linethickness{2mm} 
    \put(27,-5){\line(0,1){16}}
  \end{picture}
  \right|^2 + \cdots\\[10pt] \label{full_boltz}
  &\equiv\; - \mathcal{C}^{2\leftrightarrow 2}[f_p] - \mathcal{C}^{``1\leftrightarrow2"}[f_p] + \cdots \, ,
\end{align}
where the elastic and inelastic collision terms can be written as
\begin{align}
\notag
 \mathcal{C}^{2\leftrightarrow 2}[f_p] \;=\; &\frac{1}{2}\int \frac{d^3k\, d^3p'\, d^3k'}{(2\pi)^9} \frac{|\mathcal{M}(p,k;p',k')|^2}{(2p_0)(2k_0)(2p'_0)(2k'_0)}(2\pi)^4 \delta^{(4)}(p+k-p'-k')\\
 &\qquad \times \Big\{ f_p f_k [1\pm f_{p'}][1\pm f_{k'}] - f_{p'}f_{k'}[1\pm f_p][1\pm f_k]\Big\}\,,
\end{align}
\beq
\mathcal{C}^{``1\leftrightarrow2"}[f_p] =  \int \frac{dk}{(2\pi) 8p^2}\, \gamma(p;k)\,\big\{ f_p[1\pm f_{p-k}][1+ f_{k}] - f_{p-k}f_{k}[1\pm f_p]\big\} \, .
\eeq
The leading-order processes are illustrated on the right-hand side of (\ref{full_boltz}): elastic $2\leftrightarrow2$ scattering, 
and near-collinear $1+N\leftrightarrow2+N$ particle-splitting processes. The LPM effect is taken into account
by using an effective splitting rate that sums up all interferences in the bremsstrahlung process; the function $\gamma(p;k)$ corresponds to the near-collinear amplitude including the LPM effect, the phase-space integration and all necessary symmetry factors.

Mirroring the steps in the previous section, we discuss first the consequences of this for the elastic term. In particular, we consider as an example gluon-gluon scattering: $g \left( {\bf p} \right) + g \left( {\bf k} \right) \rightarrow g \left( {\bf p'} \right) + g \left( {\bf k'} \right)$. In the vacuum case one obtains a IR-divergent rescaled squared amplitude: 
\beq
|M(p,k;p',k')|^2 = \frac{|\mathcal{M}(p,k;p',k')|^2}{(2p_0)(2k_0)(2p'_0)(2k'_0)} \sim  \frac{\alpha^2}{(q_{\perp}^2)^2}\,,
\label{M-vacuum}
\eeq
in the limit of small momentum exchange $q_{\perp}=|\mathbf{p}-\mathbf{p}'|_{\perp}$. Including the effect of the successive scatterings in the plasma effectively results ~\cite{Arnold:2002zm} in replacing (\ref{M-vacuum}) by:
\beq
|M|^2\sim \frac{\alpha^2}{q_{\perp}^2(q_{\perp}^2+m_s^2)}\,,
\eeq
where $m_s$ denotes the screening scale, $m_s^2\sim\alpha\int d^3p \,f_p/p$,
which scales as $\alpha T^2$ once thermal equilibrium has been established. With the elastic scattering rate given by
\begin{align}
\Gamma_{\rm elas}\; &\sim\; \int d^3p\,d^3p'\,d^3k'\,|M|^2\,\delta^{(4)}(p+k-p'-k')\, f_p[1\pm f_{p'}] [1\pm f_{k'}]
\end{align}
one then finds that, for $q_{\perp}\ll |\mathbf{p}|$ \cite{Arnold:2002zm},
\begin{align}
\Gamma_{\rm elas} \;&\sim\; \int d^2q_{\perp} \frac{\alpha^2}{q_{\perp}^2(q_{\perp}^2+m_s^2)} \int d^3p\,f_p)[1\pm f_p][1\pm f_k]\\
&\sim\; \frac{\alpha^2}{m_s^2} \int d^3p\,f_p[1\pm f_p][1\pm f_k]\,.
\end{align}
Even assuming that  $1\pm f_p \sim \mathcal{O}(1)$, we find a rate  $\Gamma_{\rm elas}\sim \alpha \, p \sim \alpha \, m$, which is much larger than $H\simeq 2\Gamma_{\phi}/3v$ long before the end of reheating, for any decay rate $\Gamma_{\phi}/m\ll \alpha$. As we see below, number-increasing processes  also become effective, leading to $ f_p \gg 1$ (and thus to a even greater elastic cross section) in the bosonic case. Even though elastic processes do not increase the number of quanta, they can effectively redistribute energies towards  kinetic equilibrium.

Let us now discuss number-increasing processes. We focus the discussion on the production of ``hard'' gauge bosons, namely of a gauge boson that carries a non-negligible fraction of the energy $E$ of the primary particle. If such processes are efficient, their combined effect thermalizes the decay products by increasing the total number of quanta (pushing the ratio (\ref{ratio}) towards unity) and by decreasing the average energy per quantum. Namely, a hard primary loses energy by emitting a gauge boson with comparable, but softer, momentum $E$, which, during a time comparable to its creation time $t_{\gamma}$, splits in turn into two gauge bosons with comparable momenta (the so-called {\em hard branching}). The products of this branching quickly cascade further, giving their energy to the thermal bath \cite{Baier:2000sb}. As discussed in  \cite{Arnold:2002ja, Arnold:2001ms}, the typical time $t_\gamma$ needed for a  near-collinear emission of a hard gauge boson can be estimated as  
\beq
t_{\gamma} 
\sim  
\sqrt{\frac{\tau E}{ q_{\perp}^2}}\,,
\label{t-gamma}
\eeq
where $\tau$ is the mean time between collisions between the primary and the plasma; this expression is valid for $t_{\gamma}\lesssim E/q_{\perp}^2$  \cite{Arnold:2002zm} \footnote{For the creation of a  hard boson, this condition is equivalent to $\Gamma_{\phi}\lesssim (\frac{m^2}{M_P})v^{1/2}\simeq 10^{5}(\frac{m^3}{M_P^2})v^{1/2}$.}.
%
%
We now specialize to the hard particles with $E\sim m$ produced in inflaton decay with distribution function (\ref{f_hard}).
In the non-equilibrium case, the mean free time between small-angle scatterings, with momentum transfer of order 
$q_{\perp}\sim m_s$, scales as~\cite{Arnold:2002ja} $\tau^{-1} \sim \frac{\alpha^2}{m_s^2} \, n_p$. To estimate when the number-increasing scatterings become effective, we use the expression (\ref{ni-early}) for the number density $n_i$ of the hard decay products, and 
\beq
m_s^2\; 
\sim \; \alpha \frac{\Gamma_{\phi}^2M_P^2}{m^2}\,v^{-1}\,,
\eeq 
for the screening mass scale. This results in the following time-dependent production rate for a hard photon or gluon:
\beq
\frac{\Gamma_{\gamma,\,{\rm hard}}}{H} \sim \alpha \left(\frac{M_P}{m}\right)v^{1/2}\,,
\eeq
which is $\gg 1$ much earlier than the end of reheating. Therefore, complete thermalization can take place long before the start of the radiation-dominated era. It is worth noting that the hard gauge boson emission time-scale (and therefore the thermalization time-scale) is in general $v_{\rm hard} > v_{\rm max}$, which implies that the maximum temperature of the relativistic plasma during the reheating era is model-dependent and smaller than the maximum temperature $T_{\rm max}$ (\ref{Tmax}) assuming instantaneous thermalization.

A more detailed account of the thermalization process, tracking the phase-space distributions of the hard and soft sectors, 
can be found in~\cite{Harigaya:2013vwa,Mukaida:2015ria}. The thermalization time-scale found therein is parametrically of the same order of magnitude or smaller than the scale of hard-boson emission $v_{\rm hard}$ discussed above for a Planck-suppressed decay rate.

\section{Non-Thermal Gravitino Production}
\label{sec:nonthermal}

So far we have focused on the production of gravitinos by scattering
processes in the thermally-equilibrated plasma. In this Section we study two different mechanisms of non-thermal gravitino production during reheating. Specifically, in Subsection
\ref{sub:hard} we study the gravitinos produced by the hard inflaton decay products, before they thermalize, and in Subsection 
\ref{sub:decay} we study the amount of gravitinos produced by inflaton decays.

\subsection{Gravitino production from hard inflaton decay products}
\label{sub:hard}

In the earliest stages of reheating, before thermalization takes place,   some gravitinos  would have been produced by the scattering of the hard decay products with momenta $p\sim m$ and distribution function (\ref{f_hard}). If the number density of hard primaries were sufficiently large, the production rate would be enhanced relative to the thermal one. In this subsection we study whether this 
could increase significantly the final yield $Y_{3/2}$. 

For definiteness, we consider the specific scenario in which the inflaton field decays predominantly into gauge bosons, $\phi\rightarrow gg$:
such a scenario is possible in no-scale supergravity models of inflation with a non-trivial gauge kinetic function \cite{EGNO5,ekoty,Kallosh:2011qk}. We can then compute the non-thermal gravitino production via the channel $g^a(\bk_{1})\,+\,g^b(\bk_{2})\,\rightarrow\, \psi_{3/2}(\bp_{1})\,+\, \tilde{g}^c(\bp_{2})$, with scattering amplitude~\cite{bbb}~\footnote{The corresponding cross section for this process is infrared-finite, which allows a simple analysis.}
\beq
|\mathcal{M}|^2 = \frac{4g^2}{M_P^2}|f^{abc}|^2\left(1+\frac{m_{1/2}^2}{3m_{3/2}^2}\right)\left(s+2t + 2\frac{t^2}{s}\right)\,,
\eeq
where $t,s$ are Mandelstam variables. The Boltzmann transport equation for the gravitino distribution function is then given by
\beq\label{bltz32}
\begin{aligned}
\frac{\partial f_{p_1}}{\partial t} - Hp_1\frac{\partial f_{p_1}}{\partial p_1} &= \frac{1}{2(2p_1)} \int \frac{d^3 \bk_1}{(2\pi)^32k_1} \frac{d^3 \bk_2}{(2\pi)^32k_2} \frac{d^3 \bp_2}{(2\pi)^32p_2}\, (2\pi)^4 \delta^{(4)}(k_1+k_2-p_1-p_2)\,|\mathcal{M}|^2\\
&\qquad\qquad \times \big\{ f_{k_1}f_{k_2}[1-f_{p_2}][1-f_{p_1}] - f_{p_2}f_{p_1}[1+f_{k_1}][1+f_{k_2}] \big\}\,.
\end{aligned}
\eeq
The distribution functions $f_{k_{1,2}}$ for the gauge bosons may be approximated by (\ref{f_hard}); in the limit $\Gamma_{\phi}/m\ll 1$, they are $f_k<1$. In this approximation, we integrate the creation term of the Boltzmann equation with respect to the gravitino momentum, obtaining
\beq
\begin{aligned}\label{bltzfr}
\frac{d n_{3/2}}{d t} + 3 H n_{3/2} &\simeq  \frac{g^2|f^{abc}|^2\Gamma_{\phi}^2M_P^2}{\pi^3 (mt)^2} \left(1+\frac{m_{1/2}^2}{3m_{3/2}^2}\right) \\
&\qquad \times \int dx\,dy\,dz\ G(x,y,z)x^{-3/2}(y+z-x)^{-3/2}\, \Omega(x,y,z;t)\,,
\end{aligned}
\eeq
where the function $G(x,y,z)$ is defined as
\beq
\begin{aligned}
G(x,y,z) &\equiv  \frac{\left(x +z -\left| x-z\right| \right)}{\left(y+z\right) \left| x-z\right| } \Big[ \left(x-z\right)^2 \left( z \left(2 x-z\right) -y^2-2 y z\right)  \\
&\qquad + \left| x-z\right| \left(z \left(y+z\right){}^2+2 x^2z+\left(y^2-2 yz -z^2\right) x\right)  \Big] \,,
\end{aligned}
\eeq
and $\Omega(x,y,z;t)$ parametrizes the time-dependent integration limits, 
\begin{align}
\Omega(x,y,z;t) &= \theta(y)\, \theta(z)\, \theta\left(\tfrac{1}{2}-x\right)\, \theta\left(\tfrac{1}{2}-(y+z-x)\right)\notag \\
& \qquad \times \theta\left(x - \tfrac{b}{2}(mt)^{-2/3}\right)\, \theta\left((y+z-x) - \tfrac{b}{2}(mt)^{-2/3}\right)\,,
\end{align}
where $b= \left(\frac{3}{4}\frac{\rho_{\rm end}}{m^2M_P^2}\right)^{-1/3}$. Due to the complicated time dependence of the collision term in (\ref{bltzfr}), it is useful to consider a particular time in order to compare it to the thermal collision term. The right-hand side of (\ref{bltzfr}) can be written as 
\beq\label{cnonth}
\mathcal{C}_{\rm non-thermal} \equiv \frac{g^2|f^{abc}|^2\Gamma_{\phi}^2M_P^2}{\pi^3 } \left(1+\frac{m_{1/2}^2}{3m_{3/2}^2}\right) F(t)\,,
\eeq
where the function
\beq
F(t) \equiv (mt)^{-2} \int dx\,dy\,dz\ G(x,y,z)x^{-3/2}(y+z-x)^{-3/2}\, \Omega\left(x,y,z;t\right)
\eeq
is maximized at $mt\simeq 5.62$ with value $F\simeq 3.1\times10^{-3}$, after which it is monotonically decreasing. Evaluating then at $mt \simeq 5.62$, close to the maximum temperature during reheating $t_{\max}\sim m$ (\ref{vmax}), we find the following relation between the thermal collision term $\mathcal{C}_{\rm thermal}$, given by the right-hand side of (\ref{grav_boltz}), and the non-thermal collision term $\mathcal{C}_{\rm non-thermal}$ defined by (\ref{cnonth}),
\beq
\frac{\mathcal{C}_{\rm non-thermal}}{\mathcal{C}_{\rm thermal}}\Bigg|_{mt \simeq 5.62} \simeq \frac{10^{-4}\,g_3^2|f^{abc}|^2\Gamma_{\phi}^2M_P^2}{0.007\,c_3g_3^2\ln(k_3/g_3)T_{\max}^6/M_P^2} \sim \frac{160}{\ln(k_3/g_3)} \left(\frac{\Gamma_{\phi}}{m}\right)^{1/2}\left(\frac{M_P}{m}\right) \,,
\label{noneq3}
\eeq
where we have considered only the dominant $SU(3)$ component and disregarded $m_{1/2}$-dependent terms for simplicity. Equation (\ref{noneq3}) indicates that, 
for an inflaton decay rate $\Gamma_{\phi} \gtrsim 10^{-4}(m/M_P)^2m\sim 10^{-14} m$, the instantaneous rate for
direct production from the hard decay products is larger than the thermal production rate.
However, this is true only before thermalization is complete, i.e., $v<v_{\rm th}$, 
after which the distribution functions have their thermal forms. After thermal equilibrium is achieved, the non-thermally-produced abundance is rapidly diluted by the growing entropy density, resulting in a final abundance that is virtually indistinguishable from the abundance that would be produced in the instantaneous-thermalization case. We have verified these results by tracking numerically the gravitino yield for a decay rate $\Gamma_{\phi} = 10^{-10}m$ including non-thermal production.



\subsection{Gravitino production by inflaton decays}
\label{sub:decay}

Gravitinos may also be created by direct inflaton decay. Let us denote by $B_{3/2}$ the branching ratio of the decay to gravitinos; we implicitly assume that the number of gravitinos produced per inflaton quanta is also factored into $B_{3/2}$. Given that the number of quanta that have decayed at a time $t$ is given by (\ref{nit}), this non-thermally produced gravitino population during reheating evolves as
\beq
n_{3/2} = B_{3/2}\frac{\rho_{\phi}}{m}(e^{v}-1)\,.
\eeq
Since thermalization occurs rapidly, with $v_{\rm th}\ll 1$, we can write the ratio $n_{3/2}/n_{\rm rad}$, with the instantaneous temperature given by (\ref{inst_tempe}), as
\begin{align}
Y_{3/2,{\rm from \, decay}} (v)&= \frac{\pi^2 B_{3/2}}{\zeta(3) m }\left(\frac{g\pi^2}{30}\right)^{3/4}\frac{\rho_{\phi}}{\rho_{\gamma}^{3/4}}(e^{v}-1) \notag\\ \label{ydecay0}
&=\frac{\pi^2 B_{3/2}}{\zeta(3) m }\left(\frac{g\pi^2}{30}\right)^{3/4}\rho^{1/4}\Omega_{\gamma}^{-3/4}(1-\Omega_{\gamma})(e^{v}-1)\,.
\end{align}
The right-hand side of (\ref{ydecay0}) can be evaluated after reheating has ended by noting that, during the radiation-dominated era, $\rho\approx \rho_{\gamma} \approx \frac{3}{4}(\Gamma_{\phi}M_P)^2 v^{-2}$, and combining (\ref{rhophiex}) and (\ref{wgen}) with  $w=1/3$ for the late-time solution for $\rho_{\phi}$, 
\beq
1-\Omega_{\gamma} = \frac{\rho_{\phi}}{\rho} \simeq \frac{\frac{4}{3}(\Gamma_{\phi}M_P)^2 v^{-3/2}e^{-v}}{\frac{3}{4}(\Gamma_{\phi}M_P)^2 v^{-2}} = \frac{16}{9}v^{1/2}e^{-v}\,.
\eeq
After inclusion of the dilution factor $g(T)/g_{\rm reh}$, this leads to
\begin{align}
Y_{3/2,{\rm from \, decay}} (T\ll T_{\rm reh}) &= \frac{\pi^2 B_{3/2}}{\zeta(3)  }\left(\frac{4}{3}\right)^{7/4}\frac{g(T)}{g_{\rm reh}}\left(\frac{g_{\rm reh}\pi^2}{30}\right)^{3/4}\frac{\sqrt{\Gamma_{\phi}M_P}}{m}(1-e^{-v})\\\label{ydecay1}
&\simeq 5.9\,B_{3/2}\,\frac{\sqrt{\Gamma_{\phi}M_P}}{m}\,.
\end{align}
Comparing (\ref{ydecay1}) to the thermally-produced yield (\ref{y_final}), one finds the ratio
\beq\label{ratioofY}
\frac{Y_{3/2,{\rm from \, decay}}}{Y_{3/2,{\rm thermal}}} \simeq 1.6\times 10^3 B_{3/2} \left(\frac{M_P}{m}\right)\,.
\eeq
Hence the direct decay result would dominate if $B_{3/2} \gtrsim 10^{-8}$ for $m \simeq 10^{-5} M_P$.

This  branching ratio is model-dependent, and we focus our attention in the following on no-scale models. 
In \cite{EGNO4} we studied the decay of an untwisted sneutrino inflaton with Starobinsky potential, finding that the dominant decay rate in the presence of the Yukawa-like term $W\supset y_{\nu}H_u L \phi$ is that into the matter-Higgs channels
\beq\label{phi_decay}
\Gamma(\phi\rightarrow H_u^0\tilde{\nu},H_u^{+}\tilde{f}_L) + \Gamma(\phi\rightarrow \tilde{H}_u^0\nu,\tilde{H}_u^{+}f_L) = m\frac{|y_{\nu}|^2}{8\pi}\,,
\eeq
while the decay into gravitinos occurs with rate
\beq
\Gamma(\phi\rightarrow \psi_{3/2}\nu) =v^2\sin^2\beta\, m\frac{|y_{\nu}|^2}{32\pi M_P^2}\,,
\eeq
where in this context $v$ denotes the Higgs vev. The decay of the inflaton to a gravitino and an inflatino may also be possible if it is kinematically allowed, with a rate
\beq
\Gamma(\phi\rightarrow \psi_{3/2}\tilde{\phi}) \sim \left(\frac{m_{3/2}}{m}\right)^{2}\frac{17 m^3}{48\pi M_P^2}\,.
\eeq
The factor $(m_{3/2}/m)^2$ represents a suppression due to the near degeneracy of the inflaton and inflatino~\cite{Nilles:2001my}. The corresponding branching ratios are 
negligible: $B_{3/2}\sim (10^{-33},10^{-27}|y_{\nu}|^{-2})$ for $m_{3/2}\sim 100$\,TeV.  

An additional channel for inflaton decay into gravitinos may arise from a
superpotential term coupling the inflaton $\phi$ to the volume modulus $T$ responsible for supersymmetry breaking, of the form 
\begin{equation}
W \supset \zeta(T-1/2)^2\phi \,. 
\label{W-extra}
\end{equation}
Such terms do not spoil the inflationary potential, and may lead to a large rate for decays into gravitinos:
\beq
\Gamma(\phi\rightarrow \psi_{3/2}\psi_{3/2}) = m\frac{|\zeta|^2}{72\pi}\,.
\eeq
Assuming that the dominant channels correspond to the matter-Higgs decays (\ref{phi_decay}), and accounting for the two gravitinos that are produced in each decay, we have 
\beq
B_{3/2} = \frac{2 \vert \zeta \vert^2}{9 \, \vert y_\nu \vert^2} \;, 
\label{directratio}
\eeq
which can be sizeable. 

In the case of a volume modulus inflaton $T$, the decay rates were also evaluated in \cite{EGNO4}. The dominant channels correspond to the three-body decays 
\beq
\Gamma(T\rightarrow H_u^0 t_L \bar{t}_R,\, \tilde{t}_L\tilde{H}_u^0\bar{t}_R,\, \bar{\tilde{t}}_Rt_L \tilde{H}_u^0) = (2n_t + n_H -3)^2\frac{|y_t|^2 m^3}{12(8\pi)^3 M_P^2}\,,
\eeq
where $y_t$ denotes the top Yukawa coupling, and $n_{t,H}$ are integer modular weights. The decay to gravitinos is in this case 
\beq
\Gamma(T\rightarrow \psi_{3/2}\psi_{3/2}) \sim 10^{-3}  \left(\frac{m_{3/2}}{m}\right)^2\frac{m^3}{M_P^2}
\eeq 
up to factors at most $\mathcal{O}(1)$ that are dependent on the details of the supersymmetry-breaking sector. In this case the branching ratio into gravitinos is negligibly small.

\section{Implications for Supersymmetric Inflationary Models}
\label{sec:decay}

We now consider the implications of our results for supersymmetric models of inflation.
Our best estimate of post-inflationary gravitino production corresponds to the estimate (\ref{y_final}).
We assume that the gravitino is not the LSP, but that it is heavy enough to decay into MSSM particles,
and confront (\ref{y_final}) with constraints from Big-Bang nucleosynthesis (BBN) \cite{bbn,ceflos175,Kawasaki:1994af,stef} and
the relic cold dark matter density.

Standard BBN calculations are in good agreement with the measured light-element
abundances, with the apparent exception of Lithium \cite{cfo5}. One may use this agreement to set an upper bound
on the gravitino abundance, or one may postulate that the gravitino abundance saturates the upper bound,
in which case gravitino decays may mitigate the cosmological Lithium problem. Studies in a variety of supersymmetric
models compatible with LHC and other constraints yielded~\cite{ceflos175}
\begin{equation}
\zeta_{3/2} \; \equiv \; m_{3/2} \frac{n_{3/2}}{n_\gamma} \; = \; \frac{m_{3/2}}{2} Y_{3/2} \; \lesssim 10^{-11}\,{\rm GeV} \quad
{\rm for} \quad m_{3/2} \; \sim \; 3~{\rm TeV} \, ,
\label{zeta}
\end{equation}
rising to $\zeta_{3/2} \lesssim 10^{-8}$ GeV for $m_{3/2} \sim 6$~TeV. Combining (\ref{ytreh2}) and (\ref{zeta}),
we find 
\begin{equation}
\Gamma_\phi \; \lesssim\;  \left(1+0.56\,\frac{m_{1/2}^2}{m_{3/2}^2}\right)^{-2}
\begin{cases}
8.3 \times 10^{-6}~{\rm GeV} & {\rm for} \quad m_{3/2} = 3~{\rm TeV} \, ,\\[10pt]
2.1~{\rm GeV} & {\rm for} \quad m_{3/2} = 6~{\rm TeV} \, .
\end{cases}
\label{Gammabounds}
\end{equation}
In the case of two-body inflaton decay via a superpotential coupling $y$,
one has 
$\Gamma_\phi = \vert y \vert^2 m/8 \pi$. Assuming $m \simeq 10^{-5}M_P$, the bounds (\ref{Gammabounds})
correspond to  
\begin{equation}
\vert y \vert \; \lesssim \;  \left(1+0.56\,\frac{m_{1/2}^2}{m_{3/2}^2}\right)^{-1}
\begin{cases}
2.9 \times 10^{-9} \quad {\rm for} \quad m_{3/2} = 3~{\rm TeV} \, ,\\[10pt]
1.5 \times 10^{-6} \quad {\rm for} \quad m_{3/2} = 6~{\rm TeV} \, .
\end{cases}
\label{ybounds}
\end{equation}
These bounds would be weakened for larger values of $m_{3/2}$, disappearing altogether
if the gravitino is sufficiently heavy to decay before BBN.

However, there is another bound on $Y_{3/2}$ that applies even in this case,
which comes from the contribution to the cold dark matter density from supersymmetric relic
dark matter particles with mass $m_{\rm LSP}$ produced as the end-products of inflaton decay~\cite{EGO}:
\begin{equation}
Y_{3/2} \; < \;  \frac{2\,\Omega_{\rm cold}\,\rho_c}{m_{\rm LSP}\,n_\gamma} \, ,
\label{DMbound}
\end{equation}
where $\rho_c$ is the closure density and the factor of 2 is present because we have
defined $Y_{3/2}$ in terms of $n_{\rm rad} = n_\gamma/2$.
Using $\Omega_{\rm cold} h^2 = 0.120$ and $\rho_c = 1.054\times 10^{-5}h^2\,$cm$^{-3}$GeV, one finds
\begin{equation}
Y_{3/2} \; < \; 6.16 \times 10^{-9} \left( \frac{{\rm GeV}}{m_{\rm LSP}} \right) \, .
\label{YDMbound}
\end{equation}
Using (\ref{ytreh2}), this bound corresponds to 
\begin{equation}
\vert y \vert \; < \; 2.7 \times 10^{-5} \left(1+0.56\,\frac{m_{1/2}^2}{m_{3/2}^2}\right)^{-1}\left( \frac{100 {\rm GeV}}{m_{\rm LSP}} \right) \, ,
\label{ymLSP}
\end{equation}
and we see that $\vert y \vert \lesssim 10^{-5}$ for plausible LSP masses in the range of a few hundred GeV,
as assumed in~\cite{EGNO5}.

It was calculated in~\cite{EGNO5}  that $\vert y \vert \lesssim 10^{-5}$ corresponds, for Starobinsky-like
inflationary models, to a number of inflationary e-folds $N_* \lesssim 52$. This can to be compared
with the 68\% lower limit $N_* \gtrsim 50$ from the Planck 2015 constraint on the tilt of the scalar perturbation
spectrum. According to the calculations in~\cite{EGNO5}, the stronger BBN bound in (\ref{ybounds}) 
for $m_{3/2} = 3$~TeV would correspond to $N_* \lesssim 49$, outside the Planck 2015 68\% CL range
for Starobinsky-like models of inflation, though within the 95\% CL range $N_* \gtrsim 44$. Supersymmetric Starobinsky-like models
are clearly coming under pressure.

We conclude this Section with a discussion of non-thermally-produced gravitinos. The bound (\ref{ymLSP}) assumes 
that all gravitinos are produced by scatterings in the relativistic plasma produced during the decay of the inflaton. 
In Section~\ref{sec:nonthermal} we also computed the amount of gravitinos produced by the hard inflaton decay products (before they thermalize), 
and those produced directly in inflaton decays. While the former effect is negligible, the latter is model-dependent and gives an additional gravitino population with abundance given by (\ref{ratioofY}).
%
%
In this case the bound (\ref{YDMbound}) from the dark matter abundance on $Y_{3/2,{\rm from \, decay}}$, given by (\ref{ydecay1}), translates into 
\beq\label{B32bound}
B_{3/2}|y| < 5.2\times 10^{-11} \left(\frac{m}{M_P}\right)^{1/2}\left( \frac{100 {\rm GeV}}{m_{\rm LSP}} \right) \simeq 1.7\times 10^{-13} \left( \frac{100 {\rm GeV}}{m_{\rm LSP}} \right) \, .
\eeq
In this relations $ B_{3/2} $ is the branching ratio of the inflaton decays to gravitinos, which is model-dependent. For example, as discussed in 
Subsection (\ref{sub:decay}), in the case of an untwisted sneutrino inflaton with Starobinsky potential studied in  \cite{EGNO4}, 
a significant branching ratio can be induced by  superpotential terms that couple the inflaton $\phi$ to the volume modulus $T$
responsible for supersymmetry breaking, of the form $W\supset \zeta(T-1/2)^2\phi$. In this case,  assuming that the dominant channels 
correspond to the matter-Higgs decays (\ref{phi_decay}), the bound (\ref{B32bound}) translates into the stringent upper limit
\beq
|\zeta| < 8.7\times10^{-7}\,|y|^{1/2}\left( \frac{100 {\rm GeV}}{m_{\rm LSP}} \right)^{1/2} < 4.5\times 10^{-9} \left(1+0.56\,\frac{m_{1/2}^2}{m_{3/2}^2}\right)^{-1/2}  \left( \frac{100 {\rm GeV}}{m_{\rm LSP}} \right)\,.
\eeq

\section{Summary}

In this paper we have revisited gravitino production in supersymmetric models of inflation, with applications to Starobinsky-like models. 
Our main focus has been to examine the production of gravitinos during inflaton decay, before reheating is completed. 
Under the assumption of instantaneous thermalization (but not instantaneous inflaton decay), the dilute thermal bath formed by the inflaton decay products 
reaches a maximum temperature while the bulk of the inflaton density has yet to decay. This produces a temporary large abundance of gravitinos. 
However, these gravitinos are diluted away as the bulk of the inflaton density subsequently decays, so that naive estimates of the gravitino production 
based on instantaneous inflaton decay turn out to be rather accurate~\cite{Giudice:1999am}. 

In this work we extended previous studies by providing some semi-analytic and analytic results for the final abundance. 
We started in Section \ref{sec2} by obtaining a simple parametrization of the rate of inverse decay processes that contribute to the gravitino abundance \cite{rs}. 
While the full analytic expression of these rates is rather involved, we show that they can be parametrized accurately,
as done for the thermal scatterings in \cite{Pradler:2006qh}, for instance. Therefore, we can use the same parametrization as~\cite{Pradler:2006qh},
though with different numerical values for the parameters $c_i$ and $k_i$ so as to include the additional processes pointed out in \cite{rs}. 
This leads us to the solution (\ref{Y32-instant}) under the assumption of instantaneous inflaton decay. In this expression, we include an arbitrary parameter $c$ to describe the time
of this supposedly instantaneous decay: 
\begin{equation}
t \equiv c / \Gamma_\phi \,, 
\label{tc-summary}
\end{equation}
where $\Gamma_\phi$ is the total inflaton decay rate. The value of $c$ cannot be obtained from this simplified computation, and the only way to obtain accurately the gravitino 
abundance is to include the precise evolution of the inflaton and of the thermal number densities during the inflaton decay  \cite{rs}. 

We performed this study in Section \ref{sec:precise-reh}. Our parametrization of the gravitino production rate allows one to obtain an accurate analytic
solution for the gravitino abundance, Eq. (\ref{yasol0}). To solve this equation analytically, we use the numerical solution obtained  in \cite{EGNO5} through an iteration procedure, which was shown to be very accurate. We thus obtain the analytic solution (\ref{yc1}). To verify the accuracy of the solution, we also solved the full system of equations numerically, obtaining the result (\ref{y_final}). From this result, we could see that the analytic solution (\ref{yc1}) is accurate at the  $\lesssim 2\%$ level over the entire range $\Gamma_\phi  \lesssim  10^{-7} \, M_p$. 

The instantaneous decay approximation ignores (i) the gravitino quanta produced before $ c / \Gamma_\phi $, and (ii) the dilution of these quanta by
subsequent inflation decays. If too small a value of $c$ is chosen, an overestimate of the correct abundance results from (ii). 
On the other hand, if too large a value of $c$ is chosen,  an underestimate of the correct abundance results from (i). 
It follows that there must be some value of $c$ that provides the correct result {\emph accidentally}. The comparison between our results (\ref{Y32-instant}) and  (\ref{y_final}) 
shows that this accidental value is $c \simeq 1.2$ (in agreement with \cite{ps2}). In Fig.~\ref{fig:yfull} we compare the evolution of the gravitino abundance in the exact case vs. the instantaneous inflaton decay case, 
showing how the latter provides an overestimate of the correct result if $c=1$ is chosen. 

To obtain a complete answer on the gravitino abundance, we extended this study in two directions. In Section \ref{sec:therma}
we checked whether relaxing the assumption of instantaneous thermalization of the inflaton decay products modifies the final gravitino abundance. For Planck-suppressed inflaton decays, thermalization is mainly due to small-angle scatterings that increase the number of particles \cite{Davidson:2000er}. We verified that the maximum temperature of the thermal bath is in general lower than the result (\ref{Tmax}), since thermalization is generically delayed to later times, $v_{\rm th}>v_{\rm max}$. However, this does not affect the final gravitino abundance, as thermalization is in any case achieved well before $\Gamma_\phi \, t = {\mathcal O } \left( 1 \right)$, 
which is when most of the gravitino abundance is generated. 

As a second extension, in Section \ref{sec:nonthermal} we computed the amount of non-thermal production of gravitinos. 
We first studied the production from hard quanta generated by the inflaton decay, before thermalization takes place. 
The motivation for this analysis is that the production rate for gravitinos from scatterings increases with the energy of the incoming particles, 
and the energy of such quanta is a ${\mathcal O } \left( 1 \right)$ fraction of the inflaton mass. However, we find that the gravitinos produced by these quanta 
are diluted by the subsequent inflaton evolution and decays, and provide a negligible contribution to the final gravitino abundance. 
We also considered the gravitinos directly produced by inflaton decays. This is clearly a model-dependent study and, after a general discussion, 
we focused on no-sale supergravity models that can accommodate Starobinsky-like inflation. We find that the direct gravitino production is generally 
subdominant with respect to the thermal one, with the exception of contributions due to a specific superpotential coupling (\ref{W-extra}) of the inflaton to a
supersymmetry-breaking modulus field $T$. This does not modify the inflaton potential, and is not constrained by inflation. 

Finally, in Section \ref{sec:decay} we  studied the phenomenological implications of gravitino production. 
As is well known, gravitino production following inflation is subject to two
important constraints: late-decaying gravitinos may destroy the agreement of
standard BBN calculations with astrophysical measurements of light-element
abundances, and supersymmetric dark matter particles (LSPs) produced in
gravitino decays may have a density exceeding cosmological and astrophysical
limits. We have expressed these constraints in terms of a generic two-body
inflaton decay coupling $y$. The BBN constraint enforces the bound
$y \lesssim 2.9 \times 10^{-9}$ for $m_{3/2} = 3$~TeV, which is relaxed for
larger gravitino masses, see (\ref{ybounds}). As seen in (\ref{ymLSP}), the dark matter density constraint enforces 
$y \lesssim 10^{-5}$ for LSP masses of a few hundred GeV. Within the
context of Starobinsky-like models of inflation~\cite{EGNO5}, such as many models based
on no-scale supergravity, the former, stronger constraint would correspond to a number
of inflationary e-folds $N_*  \lesssim 49$, outside the Planck 68\% CL range
for such models, $N_*  \gtrsim 50$, whereas the latter, weaker limit would correspond to
$N_* \lesssim 52$, still compatible with the Planck 68\% CL range.

The analysis of this paper strengthens the potential connections between
inflationary cosmology and TeV-scale physics. The decays of the
inflaton into `light' particles are constrained by gravitino production as
well as CMB measurements, and gravitino production is in turn
constrained by BBN and the density of cold dark matter, by an amount
that depends on the mass of the lightest supersymmetric particle, as seen in (\ref{ymLSP}).
The constraints are becoming quite tight, implying that either some observable
signal should soon turn up, or the supersymmetric inflationary
framework discussed here may need to be rethought.
 
 \newpage
 
\section*{Acknowledgements}

The work of J.E. was supported in part by the London Centre for Terauniverse Studies
(LCTS), using funding from the European Research Council via the Advanced Investigator
Grant 267352 and from the UK STFC via the research grant ST/L000326/1.
The work of D.V.N. was supported in part by the DOE grant DE-FG02-13ER42020 and in part by the Alexander~S.~Onassis Public Benefit Foundation.
The work of M.A.G.G., K.A.O. and M.P.
was supported in part by DOE grant DE-SC0011842  at the University of
Minnesota. M.A.G.G. would like to thank C.A. Escobar for helpful discussions.


\end{document}